\documentclass[pra,twocolumn,aps,letterpaper, preprintnumbers,superscriptaddress]{revtex4-2}

\usepackage{graphicx}
\usepackage{dcolumn}
\usepackage{bm}

\usepackage{amsthm,amsmath,amssymb}
\usepackage{mathrsfs}
\usepackage{appendix}
\usepackage{amstext} 
\usepackage{url}
\usepackage{float} 
\usepackage[usenames,dvipsnames]{color}
\usepackage[colorlinks=true,citecolor=blue,urlcolor=black]{hyperref}
\usepackage{dcolumn}
\usepackage{booktabs}
\usepackage{graphicx}
\usepackage[linesnumbered,algoruled,boxed,lined]{algorithm2e} 
\usepackage{siunitx}
\usepackage{makecell}

\usepackage[dvipsnames]{xcolor}

\usepackage[utf8]{inputenc}
\usepackage[english]{babel}
\usepackage[T1]{fontenc}
\usepackage{amssymb}

\usepackage{array}
\usepackage{booktabs}
\usepackage{tikz}
\usepackage{pifont}
\usepackage{verbatim}
\usepackage{amsmath}
\usepackage{latexsym}
\usepackage{revsymb}

\usepackage{ifthen}
\usepackage{tcolorbox}
\usepackage{natbib}
\usepackage{amsfonts}
\usepackage{amsmath}
\usepackage{amssymb}
\usepackage{amsthm}
\usepackage{graphicx}
\usepackage{bm}
\usepackage{bbm}
\usepackage{epsfig,color,amssymb}
\usepackage{subfigure}
\usepackage{amsfonts}
\usepackage{amscd}
\usepackage{amsmath}
\usepackage{multirow}
\usepackage{chemarrow}
\usepackage{dcolumn}
\usepackage{bm}
\usepackage{graphicx}
\usepackage{enumerate}
\usepackage{epsfig}
\usepackage{subfigure}
\usepackage{xcolor}
\usepackage{multirow}
\usepackage{ulem}
\usepackage{braket}
\usepackage{comment}
\usepackage{enumitem}
\usepackage{amsthm}

\usepackage{stackengine}

\usepackage[justification=justified]{caption} 
\usepackage{ulem}
\renewcommand{\emph}{\textit}
\usepackage{ragged2e}
\usepackage{dsfont}
\DeclareCaptionJustification{justified}{\justifying}
\makeatletter

\newcommand{\1}{\mathds{1}}

\newcommand{\ymx}{\textcolor{black}}

\newcommand\xrowht[2][0]{\addstackgap[.5\dimexpr#2\relax]{\vphantom{#1}}}

\renewcommand{\emph}[1]{\textit{#1}}

\makeatother

\begin{document}
	
	\title{Scalable High-Rate Twin-Field Quantum Key Distribution Networks without Constraint of Probability and Intensity}
	
	\author{Yuan-Mei Xie}
	\author{Chen-Xun Weng}
	\author{Yu-Shuo Lu}
	\affiliation{National Laboratory of Solid State Microstructures, School of Physics and Collaborative Innovation Center of Advanced Microstructures, Nanjing University, Nanjing 210093, China}
	\author{Yao Fu}
	\affiliation{MatricTime Digital Technology Co. Ltd., Nanjing 211899, China}
	\author{Yang Wang}
	\affiliation{National Laboratory of Solid State Microstructures, School of Physics and Collaborative Innovation Center of Advanced Microstructures, Nanjing University, Nanjing 210093, China}
	\author{Hua-Lei Yin}\email{hlyin@nju.edu.cn}
	\affiliation{National Laboratory of Solid State Microstructures, School of Physics and Collaborative Innovation Center of Advanced Microstructures, Nanjing University, Nanjing 210093, China}
	\author{Zeng-Bing Chen}\email{zbchen@nju.edu.cn}
	\affiliation{National Laboratory of Solid State Microstructures, School of Physics and Collaborative Innovation Center of Advanced Microstructures, Nanjing University, Nanjing 210093, China}
	\affiliation{MatricTime Digital Technology Co. Ltd., Nanjing 211899, China}	

\begin{abstract}
	Implementation of a twin-field quantum key distribution network faces limitations, including the low tolerance of interference errors for phase-matching type protocols and the strict constraint regarding intensity and probability for sending-or-not-sending type protocols. Here, we propose a two-photon twin-field quantum key distribution protocol and achieve twin-field-type two-photon interference through post-matching phase-correlated single-photon interference events. We exploit the non-interference mode as the code mode to highly tolerate interference errors, and the two-photon interference naturally removes the intensity and probability constraint. Therefore, our protocol can transcend the abovementioned limitations while breaking the secret key capacity of repeaterless quantum key distribution.   Simulations show that for a four-user  networks, under which each node with fixed system parameters can dynamically switch different attenuation links, the key rates of our protocol for all six links can either exceed or approach the secret key capacity. However, the key rates of all links are lower than the key capacity when using phase-matching type protocols. Additionally, four of the links could not extract the key when using sending-or-not-sending type protocols. We anticipate that our protocol can facilitate the development of practical and efficient quantum networks.
\end{abstract}

\maketitle

\section{Introduction}
Establishing a global quantum network has become a formidable and important issue for maintaining the privacy of worldwide communications in the era of powerful quantum computers ~\cite{gidney2021factor,yan2022factoring}. Quantum key distribution (QKD)~\cite{bennett2020quantum, ekert1991quantum}, a key platform for exploring quantum networks, enables two remote users to share secret keys with information-theoretic security guaranteed by the laws of quantum physics. QKD has developed rapidly in theory and has been implemented in various experiments and networks in the past three decades ~\cite{The:Scarani:2009,weedbrook2012gaussian,xu2020secure,pirandola2020advances}. However, the non-negligible gap between ideal protocols and realistic setups underlies various security loopholes~\cite{xu2020secure}. Measurement-device-independent (MDI) QKD~\cite{lo2012measurement,braustein2012side} can eliminate these loopholes on the detection side by introducing an untrusted third party to implement entanglement swapping. Owing to excellent real-world security, ease of deployment in a star network, and sharing of expensive detectors, MDIQKD has received extensive attention and in-depth research has been conducted~ \cite{Liu2013Experimental,Rubenok:2013:Real,curty2014finite,yin2014long,ferreira2013proof,experiment2014mdi,Fu:2015:Long,azuma2015allphotonic,Making:Zhou:2016,comandar2016quantum,yin2016measurement,woodward2021gigahertz,kwek2021chip,xie2022breaking,zeng2022mode,li2022improving,gu2022experimental}. In particular, MDIQKD is regarded as an efficient and practical building block of quantum networks~\cite{tang2016measurement,pirandola2015high,Yin:2017:exp,Wang:2019:Asymmetric,fan2021measurement,zheng2021heterogeneously}.

\begin{figure}[ht!]
	\centering
	\includegraphics[width=8.6cm]{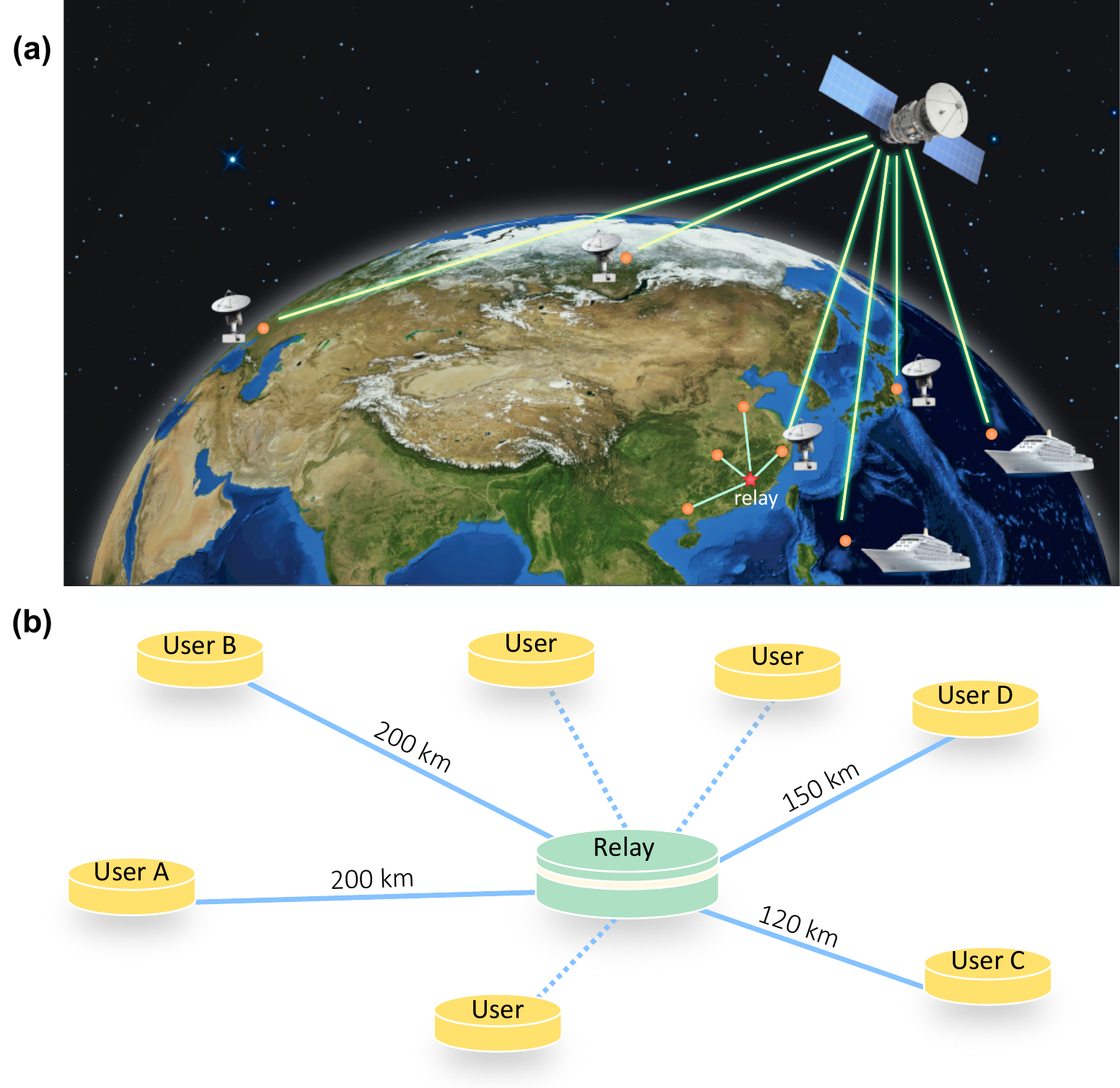}
	\caption{(a) Illustration of a satellite–ground and metropolitan-area communication network. A quantum satellite is connected to six ground stations via ground satellite links. There are two types of nodes in the quantum metropolitan-area network: user nodes (orange circles) and an untrusted relay (red star). These users are connected by a relay via fiber channels. (b) Example of a scalable QKD network setup consisting of numerous users who may freely join or leave the network. Each user node has an asymmetric channel connected to an untrusted relay, through which it can establish a QKD link to others.}\label{tptf_setup}
\end{figure}

On the other hand, optical pulses inevitably suffer from absorption and scattering in media, resulting in the key rates of MDIQKD depending heavily on quantum channel transmittance. Specifically, the extracted secure key rates of MDIQKD protocols are limited by the rate-loss scaling of repeaterless quantum channels~\cite{pirandola2009direct,takeoka2014fundamental,pirandola2017fundamental,pirandola2019end,das2021universal}. The exact limit is provided by the secret key capacity, also known as the Pirandola--Laurenza--Ottaviani--Banchi (PLOB) bound $R = -\log_2 (1-\eta)$~\cite{pirandola2017fundamental}, where $R$ and $\eta$ are the secret key rate and the channel transmittance between the two users, respectively. In the long term, this limitation can be overcome by using intermediate nodes together with quantum repeaters~\cite{duanchen2001long,sangouard2011quantum}. However, mature quantum repeaters are still far from real-life implementation because of the difficulty in constructing and reliably operating low-loss and high-clock quantum memories.

Twin-field (TF) QKD~\cite{lucamarini2018overcoming}, which exploits single-photon-type interference in the untrusted relay, has been proposed to break the PLOB bound. TFQKD maintains all the advantages of MDIQKD and greatly improves the secure key rate. Since then, many variants of TFQKD have been proposed~\cite{ma2018phase,wang2018twin,PhysRevA.98.042332,curty2019simple,cui2019twin,yin2019measurement,yin2019coherent,hu2019sending,jiang2019unconditional,maeda2019repeaterless,yin2019finite,zeng2020symmetry,curras2021tight,li2021long,yin2021twin}. These variants can be divided into two types according to the mode of extraction of the secret key. One type is the phase-matching-type protocol, including phase-matching (PM) QKD~\cite{ma2018phase}, no phase post-selection (NPP) QKD~\cite{PhysRevA.98.042332,cui2019twin,curty2019simple}, and four-phase TFQKD~\cite{wang2022twin}, \ymx{which exploits the interference mode to extract the secret key}. The other type is the single-photon-type protocol that includes sending-or-not-sending (SNS) TFQKD~\cite{wang2018twin}, which uses the non-interference mode to extract the secret key and the interference mode to estimate the phase error.
These protocols have been realized in the laboratory~\cite{minder2019experimental,wang2019beating,zhong2019proof,fang2020implementation,chen2020sending,zhong2021proof,Pittaluga2021600-km,wang2022twin,chen2022quantum} and in the field~\cite{clivati2022coherent,liu2021field,chen2021twin}. Recently, the secure transmission distance of four-phase TFQKD was experimentally demonstrated to be over 833.8 km, setting a new record for fiber-based quantum key distribution~\cite{wang2022twin}. Furthermore, the twin-field concept has been widely applied to other quantum tasks, such as device-independent quantum key distribution~\cite{xie2021overcoming}, quantum conference key agreement~\cite{grasselli2019conference,cao2021coherent,cao2021high,zhao2020phase,li2021finite}, quantum digital signatures~\cite{zhang2021twin,yin2022experimental}, and quantum secret sharing~\cite{gu2021differential,jia2021differential,gu2021secure}.

TFQKD significantly improves the key rates, extends the transmission range, and thwarts all types of detection attacks. Therefore, TFQKD is anticipated to be a major component of future communication networks~\cite{xu2020secure}. For practical quantum networks, as shown in Fig.~\ref{tptf_setup}(a), there are highly asymmetrical channels and losses at various nodes. In addition, large interference misalignment errors may occur because some nodes may be located in complex environments. Therefore, phase-matching-type protocols~\cite{ma2018phase,PhysRevA.98.042332,cui2019twin,curty2019simple,yin2019coherent,li2021long}, whose key rates are extremely sensitive to interference errors, are not considered optimal in quantum networks.
Single-photon type protocols, such as the SNS-TFQKD protocol~\cite{wang2018twin}, can tolerate large misalignment errors. With the help of actively odd-parity pairing (AOPP)~\cite{xu2020sending,jiang2020zigzag}, SNS-TFQKD greatly improves the transmission distance and the security key rate~\cite{chen2020sending,liu2021field,chen2021twin,chen2022quantum}.
However, a crucial limitation of SNS-TFQKD is that the sending probabilities and intensities of quantum states must satisfy a strict mathematical constraint (see Eq.~\eqref{mathconstra_sns}), such that the density matrix of the two-mode single-photon state in the $X$ basis is the same as that in the $Z$ basis~\cite{hu2019sending,jiang2019unconditional}. Each user has certain given system parameters in practical network links, including laser-pulse probabilities and intensities that depend on the hardware ~\cite{tang2016measurement,fan2021measurement}. Therefore, it is extremely difficult to modulate system parameters to satisfy the mathematical constraint when communication users are switched. Additionally, the light intensity cannot be precisely modulated to meet the constraint in a realistic setup, even for a fixed pair of users, resulting in an insecure key rate.

In this work, we propose a two-photon (TP) TFQKD protocol that avoids the aforementioned mathematical constraints and is robust against large misalignment error. By leveraging the idea of time multiplexing, we post-match two phase-correlated single-photon detection events to realize two-photon interference. The function of post-matching resembles a quantum repeater---the pairing for two-photon interference depends on the occurrence of  detected photons, thereby breaking the PLOB bound~\cite{pirandola2019end}. Traditional dual-rail protocols, including two-photon interference MDIQKD~\cite{lo2012measurement}, suffer from the limitation of secret key capacity~\cite{das2021universal}. Interestingly, TP-TFQKD is a dual-rail protocol that can overcome this limitation. We highlight that the key components of all QKD protocols that break the PLOB bound~\cite{duanchen2001long,azuma2015allphotonic,lucamarini2018overcoming,xie2022breaking,zeng2022mode} (including our protocol) are entanglement swapping via untrusted relays and multiplexing. Additionally, our protocol uses the non-interference mode as the code mode, which is the same as single-photon-type protocol, so it can tolerate high interference errors. Two-photon interferences in our protocol make the density matrices of the single-photon pairs always identical in the $Z$ and $X$ bases so that the security of the key rate is not affected by modulated light intensity deviations.

In a simulation of a four-user quantum network with asymmetric channels, as shown in Fig.~\ref{tptf_setup}(b), we show that TP-TFQKD enables the key rates of the five pairs of users to exceed the secret key capacity. In contrast, secret key rates cannot be generated for four pairs of users if SNS-TFQKD is adopted. When a moderate interference error occurs in the network, TP-TFQKD still enables the key rates of all six user pairs to exceed or approximate the PLOB bound, whereas the key rates of PMQKD are an order of magnitude smaller or even zero. Additionally, we derive a security proof for SNS-TFQKD when the constraint is not satisfied by using the quantum coin concept~\cite{lo2007security,koashi2009simple}. 
The transmission distance of TP-TFQKD is 200-km longer than that of SNS-TFQKD using the AOPP method if the intensities of the weak decoy states of both protocols have deviations of $3\% $. The performance of our protocol is also much better than that of PMQKD, NPPQKD, and four-phase TFQKD. Moreover, we demonstrate that the three-intensity protocol can provide higher key rates for long distances than the four-intensity protocol when finite-size effects are considered, further reducing experimental complexity. Our protocol simultaneously achieves high robustness and performance, paving the way for deployment of practical quantum networks.

\section{TP-TFQKD protocol}
\begin{figure}[t]
	\centering
	\includegraphics[width=8.6cm]{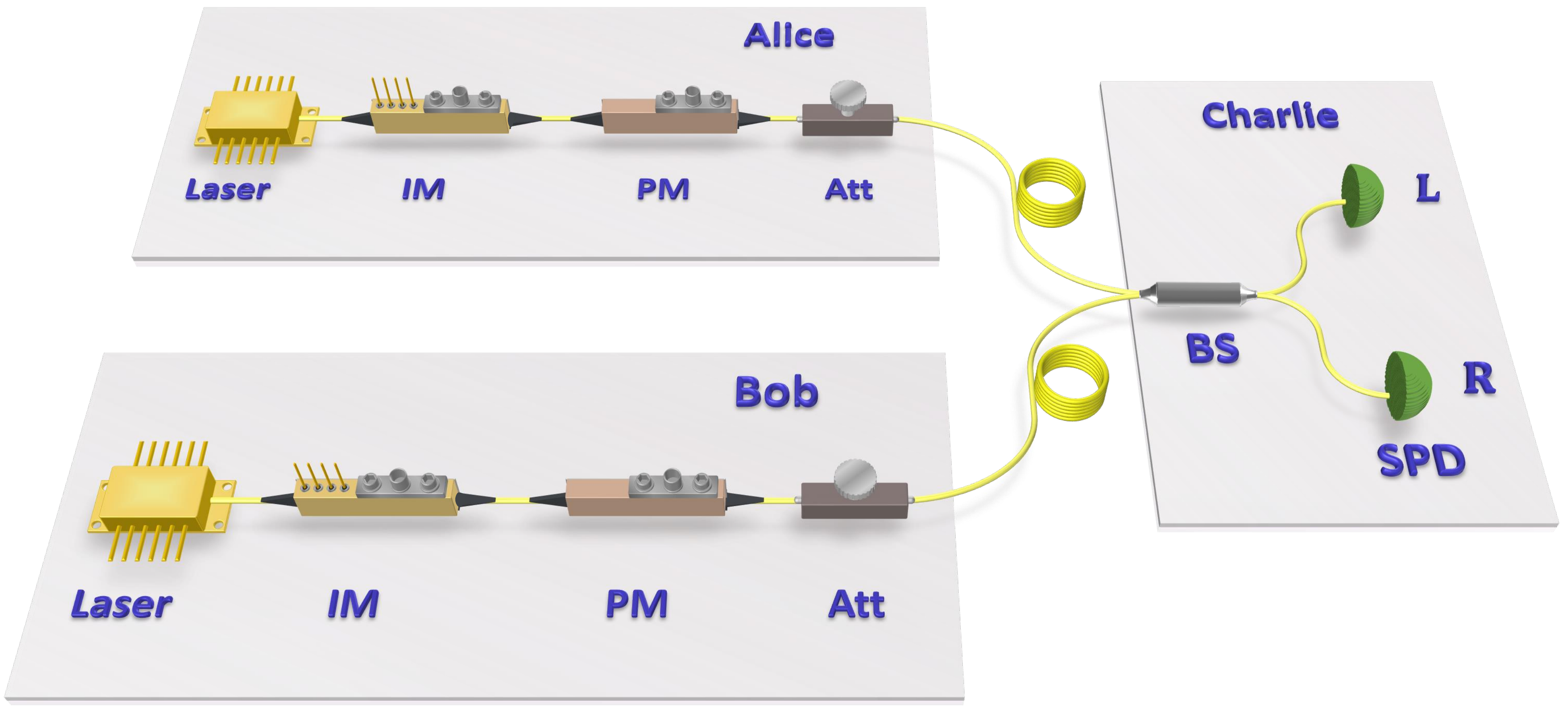}
	\caption{A schematic of the TP-TFQKD protocol setup. 
		In the station of Alice/Bob, a narrow-linewidth continuous-wave laser is utilized as the light source. Encoding is completed by an intensity modulator (IM) and a phase modulator (PM). Finally, these encoded light pulses are attenuated to the single-photon level using an attenuator (Att) and sent to Charlie.
		At the station of Charlie, the interference measurement is performed using a beam splitter (BS) and single-photon detectors (SPDs). Charlie announces detection events when only detector $\mathbf{L}$ or $\mathbf{R}$ clicks. Note that phase-locking and phase-tracking techniques are also required in our protocol.}\label{figuresetup}
\end{figure}

We illustrate the TP-TFQKD setup in Fig.~\ref{figuresetup}. Note that the TP-TFQKD setup is identical to that of TFQKD---or rather, all variants of TFQKD have the same experimental setup, and the workflows of these protocols differ only in state preparation and post-processing procedures. The experimental requirements of TP-TFQKD are also the same as those of other TFQKD protocols. A detailed description of the protocol follows.

{\it{1.~Preparation.}} Alice and Bob repeat the first two steps for $N$ rounds to obtain sufficient data. At each time bin $i\in\{1,2,\ldots,N\}$, Alice chooses a random phase, $\theta_{a}^{i}$ $\in[0,2\pi)$ and a random classical bit, $r_{a}^{i}\in\{0, 1\}$. She then prepares a weak coherent pulse $\ket{e^{\textbf{i}(\theta_{a}^{i}+r_{a}^{i}\pi)}\sqrt{k_{a}^i}}$ with probability $p_{k_a}$, where $k_a^i\in \{\mu_a,~\nu_a,~\mathbf{o}_{a},~\hat{\mathbf{o}}_{a}\}$ denote the signal, decoy, vacuum, and declare-vacuum intensities, respectively. The intensities satisfy the relation $\mu_a>\nu_a>\mathbf{o}_{a}
	=\hat{\mathbf{o}}_{a} = 0$. Similarly,
	Bob prepares a phase-randomized weak coherent pulse $\ket{e^{\textbf{i}(\theta_{b}^{i}+r_{b}^{i}\pi)}\sqrt{k_{b}^i}}$ with probability $p_{k_b}$, where $k_b^i\in \{\mu_b,~\nu_b,~\mathbf{o}_{b},~\hat{\mathbf{o}}_{b}\}$ denote the signal, decoy, vacuum, and declare-vacuum intensities, respectively. The intensities satisfy the relation $\mu_b>\nu_b>\mathbf{o}_{b}
	=\hat{\mathbf{o}}_{b} = 0$. Alice and Bob send the pulses to Charlie through insecure quantum channels. In addition, they send bright reference light to Charlie to measure the phase noise difference, $\phi_{ab}^i$.

{\it{2.~Measurement.}} Charlie performs an interference measurement on the two received pulses for each time bin and obtains a successful event when only one detector clicks. Charlie publicly announces whether he obtained a successful detection event and which detector clicked.

We represent a successful detection event as $\{k_a,~k_b\}$, where Alice sends the intensity $k_a$ and Bob sends $k_b$. The compressed notation $\{k_a^ik_a^{j},~ k_b^ik_b^{j}\}$ denotes the pairing of $\{k_a^i,~k_b^i\}$ and $\{k_a^j,~ k_b^j\}$, where the first label refers to time bin $i$, and the second to time bin $j$.

{\it{3.~Sifting.}} For time bins where Alice sends decoy or declare-vacuum intensity, or Bob sends decoy or declare-vacuum intensity, Alice and Bob communicate their intensities and phase information with each other via an authenticated classical channel.  The unannounced detection events, that is, $\{\mu_a,\mathbf{o}_{b}\}$, $\{\mu_a,\mu_b\}$, $\{\mathbf{o}_{a},\mu_b\}$, and $\{\mathbf{o}_{a},\mathbf{o}_{b}\}$, will be used to form data in the $Z$ basis. For these events, Alice randomly matches time bin $i$ of intensity $\mu_a$ with another time bin $j$ of intensity $\mathbf{o}_{a}$. She sets her bit value to 0 (1) if $i< j~(i>j)$ and informs Bob of serial numbers $i$ and $j$. Bob sets his bit value as 0 (1) if the intensities in the corresponding time bins are $k_b^{\min\{i,j\}}=\mu_b$~$(\mathbf{o}_{b})$ and $k_b^{\max\{i,j\}}=\mathbf{o}_{b}$~$(\mu_b)$. Bob announces the abandonment of the paired events where $k_b^i$ $=$ $k_b^j=\mathbf{o}_{b}$ or $\mu_b$, leaving paired events $\{\mu_a \mathbf{o}_{a},~ \mathbf{o}_{b}\mu_b\}$, $\{\mu_a\mathbf{o}_{a},~\mu_b\mathbf{o}_{b}\}$, $\{\mathbf{o}_{a}\mu_a,~ \mathbf{o}_{b}\mu_b\}$, and $\{\mathbf{o}_{a}\mu_a,~\mu_b\mathbf{o}_{b}\}$ in the $Z$ basis. Afterwards, Bob always flips his bits in the $Z$ basis. Thus, the events $\{\mu_a\mathbf{o}_{a},~\mu_b\mathbf{o}_{b}\}$ and $\{\mathbf{o}_{a}\mu_a,~ \mathbf{o}_{b}\mu_b\}$ will cause errors.

We define the global phase difference at time bin $i$ as $\theta^i:= \theta_a^i- \theta_b^i +\phi_{ab}^i$. Alice and Bob both keep detection events $\{\nu_a^i,~\nu_b^i\}$ only if $\theta^i~\in[-\delta,\delta]\cup[\pi-\delta,\pi+\delta]$. Here, the value $\delta$ is determined by the size of the phase slice chosen by Alice and Bob. They randomly select two retained detection events that satisfy $\left|\theta^i -\theta^j\right|\approx 0 $ or $\pi$, and then match these two events, denoted as $\{\nu_a^i\nu_a^j,\nu_b^i\nu_b^j\}$. By calculating the classical bits $r_a^i \oplus r_a^j $ and $r_b^i \oplus r_b^j $, Alice and Bob extract a bit value on the $X$ basis, respectively. Bob flips his bits when the global phase difference between two matching time bins is 0~($\pi$) and the two clicking detectors announced by Charles are different (same) in the $X$ basis. 
The events $\{\mu_a,\hat{\mathbf{o}}_b\}$, $\{\hat{\mathbf{o}}_a,\mu_b\}$, $\{o_a,\nu_b\}$, $\{\nu_a,o_b\}$,  $\{\hat{\mathbf{o}}_a,o_b\}$, and $\{\mathbf{o}_a, \hat{\mathbf{o}}_b\}$ will be used for decoy analysis, where $o_{a(b)}\in\{\mathbf{o}_{a(b)}, \hat{\mathbf{o}}_{a(b)}\}$.

{\it{4.~Parameter estimation.}} Alice and Bob use random bits from the $Z$ basis to form $n^z$-length raw key bits. The remaining bits in the $Z$ basis are used to compute the bit error rate, $E^z$. All bit values in the $X$ basis are disclosed to obtain the total number of errors. Alice and Bob use the decoy-state method~\cite{wang2005beating,lo2005decoy} to estimate  the number of 
events in the $Z$ basis when Alice emits a zero-photon state  in the two matched time bins, $s_{0\mu_b}^{z}$, the count of single-photon pairs in the $Z$ basis, $s_{11}^z$, the bit error rate in the $X$ basis, $e_{11}^x$, and the phase error rate of single-photon pairs in the $Z$ basis, $\phi_{11}^z$~(see Appendix~\ref{simu_p1} for details).

{\it{5.~Postprocessing.}} Alice and Bob distill the final keys utilizing the error correction algorithm with $\varepsilon_{\rm{cor}}$-correct and the privacy amplification algorithm with $\varepsilon_{\rm{sec}}$-secret.  Using the entropic uncertainty relation~\cite{curty2014finite}, the length of the final secret key, $\ell$, against coherent attacks with total security $\varepsilon_{\rm{TP}}=\varepsilon_{\rm{sec}} + \varepsilon_{\rm{cor}}$ can be written as
\begin{equation}
	\begin{aligned}\label{eq_keyrate_finite}
		\ell=&\underline{s}_{0\mu_b}^z+\underline{s}_{11}^z\left[1-H_2(\overline{\phi}_{11}^z)\right]-n^zfH_2(E^z) \\
		&-\log_2\frac{2}{\varepsilon_{\rm cor}}-2\log_2\frac{2}{\varepsilon'\hat{\varepsilon}}-2\log_2\frac{1}{2\varepsilon_{\rm PA}},
	\end{aligned}
\end{equation}
and $\varepsilon_{\rm sec}=2(\varepsilon'+\hat{\varepsilon}+2\varepsilon_e)+\varepsilon_\beta+\varepsilon_0+\varepsilon_1+\varepsilon_{\rm PA}$. Here, 
\begin{itemize}
	\setlength{\partopsep}{0pt}
	\setlength{\listparindent}{0em}
	\setlength{\itemsep}{0.5pt}
	\setlength{\parsep}{0pt}
	\setlength{\parskip}{0pt}
	\item [$\underline{x}$] is the lower bound of observed value $x$.
	\item [$\overline{x}$] is the upper bound of observed value $x$.
	\item [$H_2(x)$] is the binary Shannon entropy function: $H_2(x)=-x\log_2x-(1-x)\log_2(1-x)$.
	\item [ $f$ ] is the error correction efficiency. 
	\item [$\varepsilon_{\rm cor}$] is the failure probability of error verification.
	\item [$\varepsilon_{\rm PA}$] is the failure probability of privacy amplification.
	\item [ $\varepsilon'$] and $\hat{\varepsilon}$ are coefficients when using chain rules of smooth min-entropy and max-entropy, respectively.
   \item [$\varepsilon_\beta$]    is the failure probability for estimating the terms of $E_z$.
	\item [$\varepsilon_0$], $\varepsilon_1$, and $\varepsilon_e$ are the failure probabilities for estimating the terms of $s_{0\mu_b}^z$, $s_{11}^z$, and $\phi_{11}^z $, respectively. 
\end{itemize}

\section{Security proof}
Here, we introduce a virtual entanglement-based TP-TFQKD protocol. Let $\ket{10}^{ij}=\ket{1}^{i}\ket{0}^{j}$ denote the quantum state, where we have one photon in time bin $i$ and zero photons in time bin $j$.
For time bins $i$ and $j$, Alice (Bob) prepares an entangled state $\ket{\phi^+}_{Aa}=1/\sqrt{2}(\ket{+z}_A\ket{+z}_a+\ket{-z}_A\ket{-z}_a)$~($\ket{\phi^+}_{Bb}$), where $\ket{\pm z}_{A (B)}$ are two eigenvectors of the Pauli $Z$ operator denoting the qubits in the hands of Alice (Bob), and $\ket{+z}_{a (b)}=\ket{10}^{ij}$ and $\ket{-z}_{a (b)}=\ket{01}^{ij}$ are two eigenvectors of the $Z$ basis for optical modes $a~(b)$, respectively. Accordingly, the eigenvectors of the $X$ basis are $\ket{\pm x}=(\ket{+z}\pm\ket{-z})/\sqrt{2}$. The joint quantum state of Alice and Bob can be written as
\begin{equation}
	\begin{aligned}
		\ket{\Phi}_{AaBb}=\ket{\phi^+}_{Aa}&\otimes	\ket{\phi^+}_{Bb}\\
		=\frac{1}{2}(\ket{+z+z}_{AB}&\ket{+z+z}_{ab}+\ket{-z-z}_{AB}\ket{-z-z}_{ab}\\
		+\ket{+z-z}_{AB}&\ket{+z-z}_{ab}+\ket{-z+z}_{AB}\ket{-z+z}_{ab}),\\
	\end{aligned}
\end{equation}
where quantum state $\ket{+z+z}_{ab}$ represents $\ket{+z}_{a}\ket{+z}_{b}$.
Next, Alice and Bob perform nonlocal operations to preserve their quantum state of unequal qubits. For example, Alice and Bob can perform controlled NOT operations on an auxiliary qubit $\ket{+z}_{C}$ on Alice's side controlled by qubit $A$ and qubit $B$, respectively~(Bob could perform the controlled NOT nonlocally with the help of an additional entangled state~\cite{eisert2000Optimal}). Alice then measures qubit $C$ in the $Z$ basis and they post select the joint quantum state when the outcome is $-z$.
Under this condition, the joint quantum state acquired by Alice and Bob can be written as
\begin{equation}
	\begin{aligned}\label{eq6}
		\ket{\Psi}_{ABab}=&\frac{1}{\sqrt{2}}(\ket{+z-z}_{AB}\ket{+z-z}_{ab}\\
		&+\ket{-z+z}_{AB}\ket{-z+z}_{ab})\\
		=&\frac{1}{\sqrt{2}}\left(\frac{\ket{+x+x}_{AB}-\ket{-x-x}_{AB}}{\sqrt{2}}\ket{\psi^{+}}_{ab}\right.\\
		&-\left.\frac{\ket{-x+x}_{AB}-\ket{+x-x}_{AB}}{\sqrt{2}}\ket{\psi^{-}}_{ab}\right).\\
	\end{aligned}
\end{equation}
Alice and Bob hold qubits $A$ and $B$, respectively. They send optical modes $a$ and $b$, respectively, to Charlie, who performs Bell state measurements to identify the two Bell states: $\ket{\psi^+}_{ab}=(\ket{+z-z}_{ab}+\ket{-z+z}_{ab})/\sqrt{2}$ and $\ket{\psi^-}_{ab}=(\ket{+z-z}_{ab}-\ket{-z+z}_{ab})/\sqrt{2}$. The same detector clicks at time bins $i$ and $j$ indicate a projection into Bell state $\ket{\psi^+}_{ab}$. Different detector clicks at time bins $i$ and $j$ imply a projection into Bell state $\ket{\psi^-}_{ab}$. 

\begin{table}[b]
	\centering
	\caption{Simulation parameters. $\eta_{d}$ and $p_{d}$ are the detector efficiency and dark count rate, respectively. $\alpha$ is the attenuation coefficient of the fiber, $e_d^z$ is the misalignment error in the $Z$ basis, and $f$ denotes the error correction efficiency. $\epsilon$ is the failure probability considered in the error verification and finite data analysis processes.}\label{tab1}
	\begin{tabular}[b]{@{\extracolsep{13pt}}cccccc}
		\hline
		\hline
		$\eta_{d}$  & $p_{d}$      & $\alpha$ &$e_d^z$& $f$  & $\epsilon$\\
		\hline\xrowht{7pt}
		$70\%$ & $10^{-8}$     & $0.165$ dB/km & 0&$1.1$  &
		$36/24\times10^{-10}$\\
		\hline
		\hline
	\end{tabular}
\end{table}

\begin{table*}[tb]
	\centering
	\caption{Simulated secure key rates for TP-TFQKD and SNS-TFQKD with the AOPP method in the QKD network shown in Fig.\ref{tptf_setup}(b) using the parameters in Table~\ref{tab1}. The angle of misalignment in the $X$ basis of TP-TFQKD and SNS-TFQKD~(AOPP) are both set to $\sigma=5^{\circ}$ and the data size is $N=10^{11}$. Here, link A-C represents that user A communicates with user C. The sending intensities and corresponding probabilities are selected by the users to obtain the optimal key rate for links A to C and B to D. Here, the superscripts $\rm{tp}$ and $\rm{sns}$ correspond to the TP-TFQKD and SNS-TFQKD~(AOPP) protocols, respectively. When users switch to communicate with other users, these values remain unchanged.}\label{tab3}
	\begin{tabular}[b]{@{\extracolsep{7pt}}ccccccccccccccc}
		\hline
		\hline
		\xrowht{8pt}
		&\multicolumn{2}{c}{Protocol} &\multicolumn{2}{c}{A-C}  &\multicolumn{2}{c}{B-D}  &\multicolumn{2}{c}{A-D}  &\multicolumn{2}{c}{A-B} &\multicolumn{2}{c}{C-D} &\multicolumn{2}{c}{C-B}\\
		\hline\xrowht{8pt}
		&\multicolumn{2}{c}{PLOB} & \multicolumn{2}{c}{$5.300\times10^{-6}$}  &\multicolumn{2}{c}{$1.695\times10^{-6}$}  &\multicolumn{2}{c}{$1.695\times10^{-6}$}  & \multicolumn{2}{c}{$2.537\times10^{-7}$} &\multicolumn{2}{c}{$3.542\times10^{-5}$} &\multicolumn{2}{c}{$5.300\times10^{-6}$}\\
		&\multicolumn{2}{c}{TP-TFQKD} & \multicolumn{2}{c}{$8.631\times10^{-6}$}  & \multicolumn{2}{c}{$6.701\times10^{-6}$}  & \multicolumn{2}{c}{$6.063\times10^{-6}$}  & \multicolumn{2}{c}{$1.743\times10^{-6}$} &\multicolumn{2}{c}{$1.754\times10^{-5}$} &\multicolumn{2}{c}{$8.456\times10^{-6}$}\\
		&\multicolumn{2}{c}{SNS-TFQKD~(AOPP)} & \multicolumn{2}{c}{$1.111\times10^{-5}$}  &\multicolumn{2}{c}{$8.228\times10^{-6}$}  &\multicolumn{2}{c}{$ 0$}  & \multicolumn{2}{c}{$0$} &\multicolumn{2}{c}{$0$} &\multicolumn{2}{c}{$0$}\\
		\hline
		\xrowht{8pt}
		&\multicolumn{2}{c}{\multirow{2}{*}{User}}& \multicolumn{6}{c}{TP-TFQKD} & \multicolumn{6}{c}{SNS-TFQKD~(AOPP)}\\
		\cline{4-9} \cline{10-15}
		\xrowht{8pt}
		&\multicolumn{2}{c}{}    &$\mu^{\rm{tp}} $  &$\nu^{\rm{tp}} $  &$p_{\mu}^{\rm{tp}}$  &$p_{\nu}^{\rm{tp}}$ &$p_{\mathbf{o}}^{\rm{tp}}$ &$p_{\hat{\mathbf{o}}}^{\rm{tp}}$ &$\mu^{\rm{sns}}$  &$\nu^{\rm{sns}}$  &$p_{\mu}^{\rm{sns}}$  &$p_{\nu}^{\rm{sns}}$ &$p_{\mathbf{o}}^{\rm{sns}}$ &$p_{\hat{\mathbf{o}}}^{\rm{sns}}$\\
		\hline\xrowht{8pt}
		&\multicolumn{2}{c}{A} & $0.725$  & $0.100$  & $0.388$  & $0.159$ &$0.449$ &$0.004$ & $0.789$  & $0.1262$  & $0.388$  & $0.111$ &$0.498$ &$0.003$\\
		&\multicolumn{2}{c}{B} & $0.681$  &$0.098$  & $0.362$  & $0.163$ &$0.471$ &$0.004$  & $0.765$  & $0.0714$  & $0.336$  & $0.124$ &$0.537$ &$0.003$\\
		&\multicolumn{2}{c}{C} & $0.166$  &$0.005$  &$0.069$  & $0.161$ &$0.762$ &$0.008$ & $0.164$  &$0.006$  &$0.077$  & $0.108$ &$0.807$ &$0.008$\\
		&\multicolumn{2}{c}{D} & $0.250$  & $0.015$  & $0.108$  & $0.165$ &$0.716$ &$0.011$  & $0.253$  & $0.010$  & $0.119$  & $0.125$ &$0.749$ &$0.007$\\
		\hline
		\hline
	\end{tabular}
\end{table*}

After Charlie announces whether he has obtained a Bell state and which Bell state he has identified through public channels, Alice and Bob will randomly choose the $Z$ or $X$ basis to measure qubits $A$ and $B$, respectively. Alice and Bob publish the basis information through an authenticated classical channel. Bob always flips his bit value if he chooses the $Z$ basis.
Bob will flip his outcome in the $X$ basis if Charlie's announcement is $\ket{\psi^-}_{ab}$. They use the data of the $Z$ basis to generate the raw key, while the data of the $X$ basis are used to estimate the leaked information. Alice and Bob can acquire a secure key through error correction and privacy amplification.

Next, the entanglement-based protocol can be equivalently converted to the prepare-and-measurement protocol by the arguments of Shor--Preskill. For the prepare-and-measure protocol, Alice and Bob prepare the quantum states of modes $a$ and $b$ according to Eq.~\eqref{eq6}. Specifically, Alice and Bob send joint quantum states of a single-photon pair, $\ket{+z-z}_{ab}$ and $\ket{+z-z}_{ab}$, randomly if they choose the $Z$ basis. They randomly send the joint quantum states of a single-photon pair, $\ket{\psi^{+}}_{ab}$ and $\ket{\psi^{-}}_{ab}$, if they choose the $X$ basis. 
The other steps are all the same as the entanglement-based protocol, including the Bell state measurement, basis comparison, bit flipping, error correction, and privacy amplification. After Alice and Bob know the bit error rate of the state with single-photon pairs $e_{11}^x$ in the $X$ basis, the asymptotic key rate can be given by~\cite{lo1999unconditional,shor2000simple}
\begin{equation}
	\begin{aligned}
		R&=q_{11}[1-H_2(e_{11}^x)-fH_2(e_{11}^z)],\\
	\end{aligned}
\end{equation}
where $q_{11}$ and $e_{11}^z$ represent the gain and bit error rates of the single-photon pairs in the $Z$ basis, respectively.
Note that the entanglement-based protocol and prepare-and-measure protocol are virtual protocols, which are not used experimentally but to prove security in theory. In fact, Alice and Bob do not need to pre-share the nonlocal entanglement state $\ket{\psi^\pm}_{ab}$. Instead, they can send the two-mode phase-randomized coherent states $\ket{e^{i(\kappa+r_{a}^{i}\pi)}\alpha_{a}}_{a}^{i}\ket{e^{i(\varphi+r_{a}^{j}\pi)}\alpha_{a}}_{a}^{j}$ and $\ket{e^{i(\kappa+r_{b}^{i}\pi)}\alpha_{b}}_{b}^{i}\ket{e^{i(\varphi+r_{b}^{j}\pi)}\alpha_{b}}_{b}^{j}$ to Charlie in the $X$ basis. Then, they use the decoy-state method to estimate the upper bound of the phase-flip error rate of $\ket{\psi^\pm}_{ab}$~(see Appendix~\ref{secur_TP}).

\section{Performance and Discussion}
\label{sec_discussion}

We first provide an intuitive understanding of the scaling of our protocol before demonstrating its performance. In the TP-TFQKD protocol, successful detection in each time bin depends on the arrival of a single photon at the relay with a probability of $O(\sqrt{\eta})$, which is identical to the original twin-field protocol. Afterwards, the post-matching process, which is the pairwise matching of successful detection events, will only add a factor of 1/2 to the $O(\sqrt{\eta})$ scaling for the large data size $N$.  Compared with traditional two-photon interference, wherein the pairings are predetermined independently of the arrival of a photon at the relay,  twin-field-type asynchronous two-photon interference improves the scaling from $O(\eta)$ to $O(\sqrt{\eta})$.

\begin{table*}[tb]
	\centering
	\caption{Simulated secure key rates for TP-TFQKD and PMQKD in the QKD network shown in Fig.\ref{tptf_setup}(b) using parameters in Table~\ref{tab1}. The angle of misalignment in the $X$ basis of TP-TFQKD and PMQKD were both set to $\sigma=18^{\circ}$ and the data size is $N=10^{11}$. The users fix sending intensities and corresponding probability values to values that enable the A-C and B-D links to obtain the optimal key rate. Here, superscripts $\rm{tp}$ and $\rm{pm}$ correspond to TP-TFQKD and PMQKD protocols, respectively. }\label{tab4}
	\begin{tabular}[b]{@{\extracolsep{7pt}}ccccccccccccc}
		\hline
		\hline
		\xrowht{8pt}
		{Protocol} &\multicolumn{2}{c}{A-C}  &\multicolumn{2}{c}{B-D}  &\multicolumn{2}{c}{A-D}  &\multicolumn{2}{c}{A-B} &\multicolumn{2}{c}{C-D} &\multicolumn{2}{c}{C-B}\\
		\hline\xrowht{8pt}
		{PLOB} & \multicolumn{2}{c}{$5.300\times10^{-6}$}  &\multicolumn{2}{c}{$1.695\times10^{-6}$}  &\multicolumn{2}{c}{$1.695\times10^{-6}$}  & \multicolumn{2}{c}{$2.537\times10^{-7}$} &\multicolumn{2}{c}{$3.542\times10^{-5}$} &\multicolumn{2}{c}{$5.300\times10^{-6}$}\\
		{TP-TFQKD} & \multicolumn{2}{c}{$2.572\times10^{-6}$}    & \multicolumn{2}{c}{$1.945\times10^{-6}$}  & \multicolumn{2}{c}{$1.730\times10^{-6}$} & \multicolumn{2}{c}{$3.573\times10^{-7}$} &\multicolumn{2}{c}{$6.192\times10^{-6}$} &\multicolumn{2}{c}{$2.530\times10^{-6}$}\\
		{PMQKD} & \multicolumn{2}{c}{$2.840\times10^{-7}$}  & \multicolumn{2}{c}{$2.400\times10^{-7}$}  & \multicolumn{2}{c}{$2.244\times10^{-7}$}  & \multicolumn{2}{c}{$0$} &\multicolumn{2}{c}{$8.940\times10^{-7}$} &\multicolumn{2}{c}{$2.700\times10^{-7}$}\\
		\hline\specialrule{0em}{1pt}{1pt}
		\multicolumn{2}{c}{\multirow{2}{*}{User}}&   \multicolumn{6}{c}{TP-TFQKD} & \multicolumn{5}{c}{PMQKD}\\
		\cline{3-8} \cline{9-13}  \noalign{\smallskip}& &
		$\mu^{\rm{tp}} $  & $\nu^{\rm{tp}} $  & $p_{\mu}^{\rm{tp}}$  & $p_{\nu}^{\rm{tp}}$ &$p_{\mathbf{o}}^{\rm{tp}}$ &$p_{\hat{\mathbf{o}}}^{\rm{tp}}$ &$\mu^{\rm{pm}}$  &$\nu^{\rm{pm}}$  &$p_{\mu}^{\rm{pm}}$  &$p_{\nu}^{\rm{pm}}$ &$p_{o}^{\rm{pm}}$\\
		\hline\specialrule{0em}{1pt}{1pt}
		\multicolumn{2}{c}{A}&   0.720&    0.084&    0.343&    0.243&    0.409&    0.005  &0.086&    0.024&    0.605&   0.176&    0.219\\
		\multicolumn{2}{c}{B}& 0.666&    0.084&    0.320&    0.245&    0.430&    0.005  & 0.076&    0.022&    0.613&    0.169&    0.218\\
		\multicolumn{2}{c}{C}&0.168&    0.004&    0.058&    0.251&    0.677&    0.014 & 0.005&   0.002&   0.626&    0.172&    0.202 \\
		\multicolumn{2}{c}{D}& 0.266&    0.013&    0.088&    0.262&    0.638&   0.012& 0.013&    0.004&    0.609&    0.177&    0.214\\
		\hline
		\hline
	\end{tabular}
\end{table*}

\subsection{Quantum network simulation results}
To show the TP-TFQKD performance in a real-life network, we focus on the four users (A, B, C, and D) in the QKD network shown in Fig. \ref{tptf_setup}(b), where the corresponding distances to the untrusted relay are $L_A=200$~km, $L_B=200$~km, $L_C=120$~km, and $L_D=150$~km.
The experimental parameters were set to be typical values and are listed in Table~\ref{tab1}. For the finite-size regime, we set failure parameters $\varepsilon_{\rm{cor}}$, $\varepsilon'$, $\hat{\varepsilon}$, $\varepsilon_e$, $\varepsilon_\beta$, and $\varepsilon_{\rm PA}$ to be the same value: $\epsilon$. We have $\varepsilon_0+\varepsilon_1=13\epsilon$ because we use the Chernoff bound~\cite{chernoff1952measure, yin2020tight} 13 times to estimate $s_{0\mu_b}^z$, $s_{11}^z$, and $e_{11}^x$. The corresponding security bound is $\varepsilon_{\rm{TP}}=24\epsilon$. We also simulated the key rate of the SNS-TFQKD protocol using the actively odd-parity pairing method (simply AOPP hereafter) for comparison. In Ref.~\cite{jiang2020zigzag}, the security bound of the AOPP is $\varepsilon_{\rm{AOPP}}=36 \varepsilon$ in finite-size cases. For a fair comparison, we set $\varepsilon=10^{-10}$ and $\epsilon=36{\varepsilon}/24$ such that $\varepsilon_{\rm{TP}}=\varepsilon_{\rm{AOPP}}=3.6\times 10^{-9}$. Detailed equations for simulating the TP-TFQKD protocol are shown in Appendix~\ref{simu_pb}.
In a commercialized network, the sending intensities and corresponding probabilities are usually fixed as long as the hardware is set up. Here, we set the sending intensity and probability of four users, which will allow links A-C and B-D to have optimal key rates. If the AOPP protocol is utilized in other links (A-D, A-B, C-D, and C-B), the mathematical constraints required by the AOPP (given in Eq.~\ref{mathconstra_sns} in Appendix~\ref{app_sns}) may not be satisfied, resulting in completely insecure key rates. In this case, we provide quantitative bounds that connect deviation from the mathematical constraints to the security of the SNS-TFQKD protocol, thus giving a secure key rate when the constraint is violated (see Appendix~\ref{app_sns} for details).

In Table~\ref{tab3}, we list the secret key rate, $R=\ell/N$, for different pairs of users under the TP-TFQKD and AOPP protocols ~\cite{jiang2020zigzag} when the data size is $N=10^{11}$. To assess the performance of the network, we also provide the PLOB bound. For links A-C and B-D, the key rates for TP-TFQKD are only slightly lower and comparable to those of AOPP. If TP-TFQKD is to be implemented, links A-D, A-B, C-D, and C-B also have high key rates, whereas no secure key is generated with the AOPP protocol. In addition, five pairs of links among the six pairs can exceed the PLOB bound with TP-TFQKD. In view of this, our protocol is suitable not only for long-distance communication but also for scalable networks.

\begin{figure}[t!]
	\centering
	\includegraphics[width = \columnwidth]{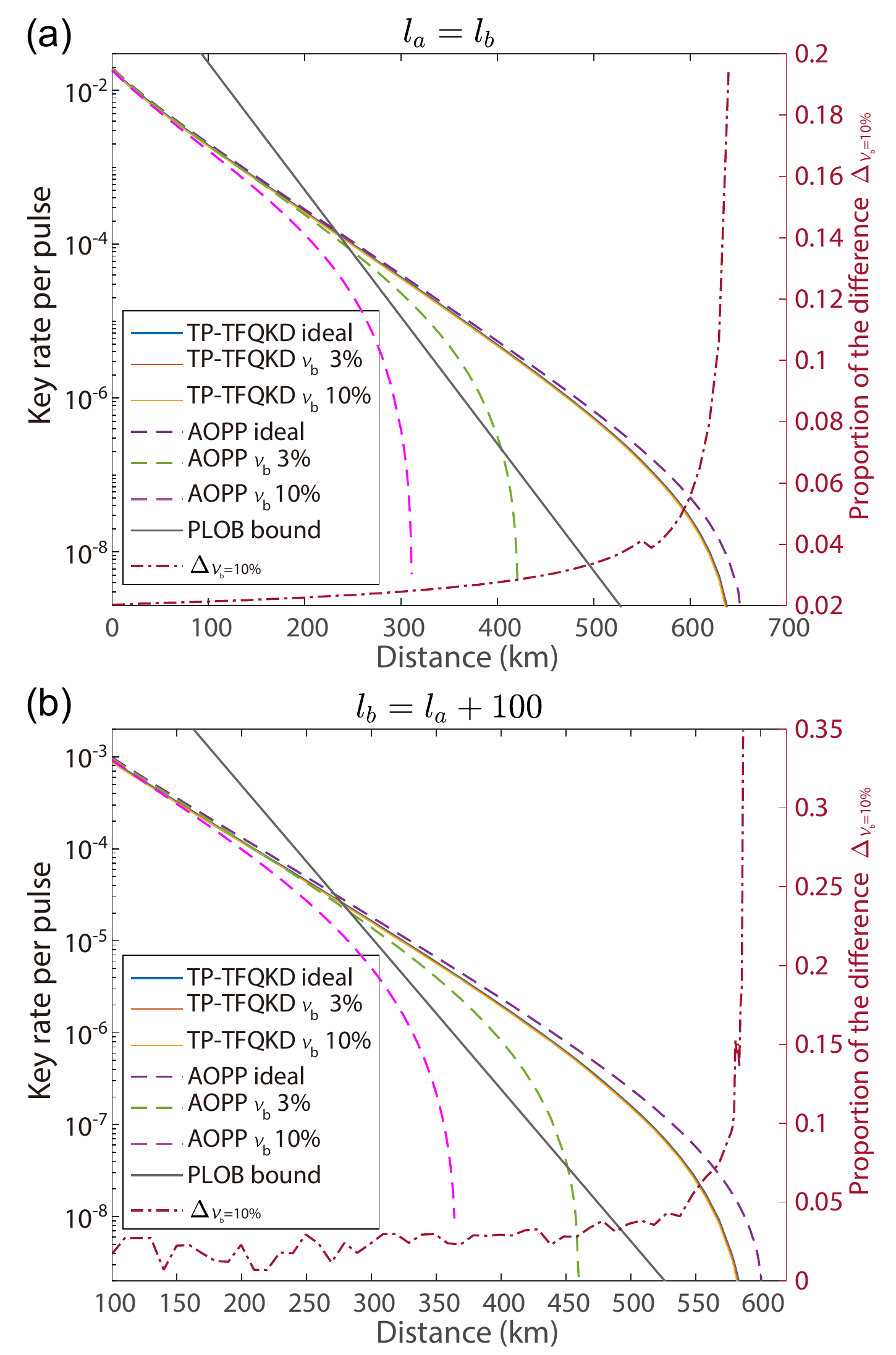}
	\caption{
		Comparison of the secret key rates of TP-TFQKD and AOPP~\cite{jiang2020zigzag} in the finite-size regime under two types of channels: (a) symmetric channel $l_a=l_b$ and (b) asymmetric channel $l_b-l_a=100$~km. The horizontal axis represents total transmission distance $l=l_a+l_b$ and the parameters listed in Table~\ref{tab1} and $N=10^{13}$ were used. The angles of misalignment in the $X$ basis of TP-TFQKD and AOPP were both set to $\sigma=5^{\circ}$ and the security bounds are $\varepsilon_{\rm{TP}}=\varepsilon_{\rm{AOPP}}=3.6\times 10^{-9}$. In each plot, the solid (dashed) lines represent the key rates of TP-TFQKD (AOPP) for different preparation deviations of decoy intensity $\nu_b$, and the solid gray line illustrates the PLOB bound~\cite{pirandola2017fundamental}. The simulations show that our protocol can achieve key rates comparable to those of the AOPP scheme in the ideal case. Additionally, TP-TFQKD obtained a significantly higher key rate than AOPP in the presence of more than $3\%$ deviation of modulated light intensity $\nu_b$. }
	\label{figureaoppun_fig}
\end{figure}

\begin{figure}[t!]
	\centering
	\includegraphics[width = \columnwidth]{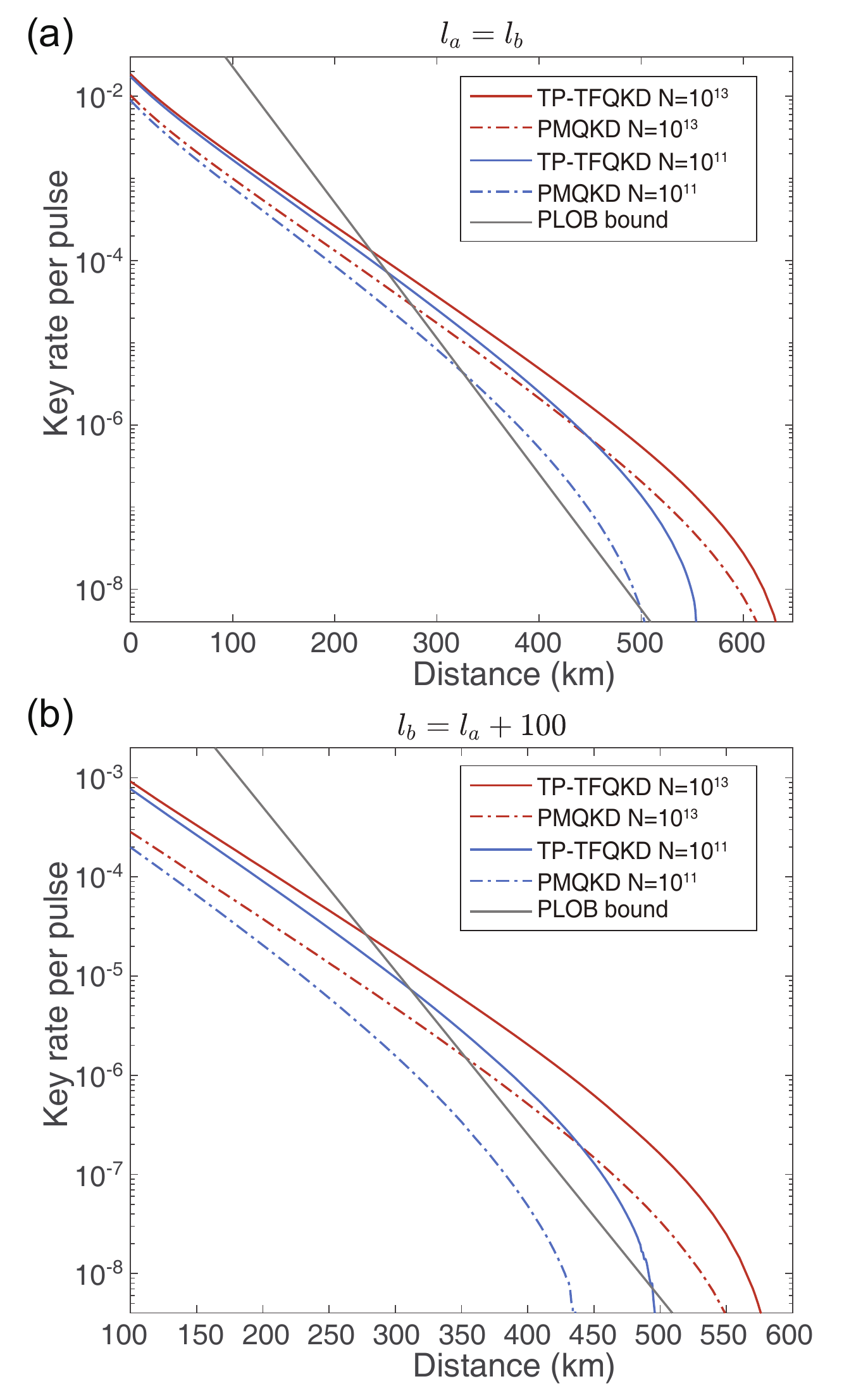}
	\caption{(a) Comparison of the secret key rates of TP-TFQKD and PMQKD~\cite{zeng2020symmetry} under two channel types: (a) symmetric, $l_a=l_b$, and (b) asymmetric, $l_b-l_a=100$~km. The horizontal axis represents the total transmission distance, $l=l_a+l_b$. The parameters listed in Table~\ref{tab1} were used and the angles of misalignment in the $X$ basis of TP-TFQKD and PMQKD were both set to $\sigma=5^{\circ}$. In the finite case, we set the failure probability of PMQKD to $10^{-10}$, and the corresponding security bound was $O(10^{-9})$. In the finite case, we simulated two groups of results where $N = 10^{11}$ and $N = 10^{13}$. The simulations show that TP-TFQKD has a higher key rate than PMQKD.
		}\label{fig_keyrate_finite_sym}
	\end{figure}

	\begin{figure}[t!]
		\centering
		\includegraphics[width = \columnwidth]{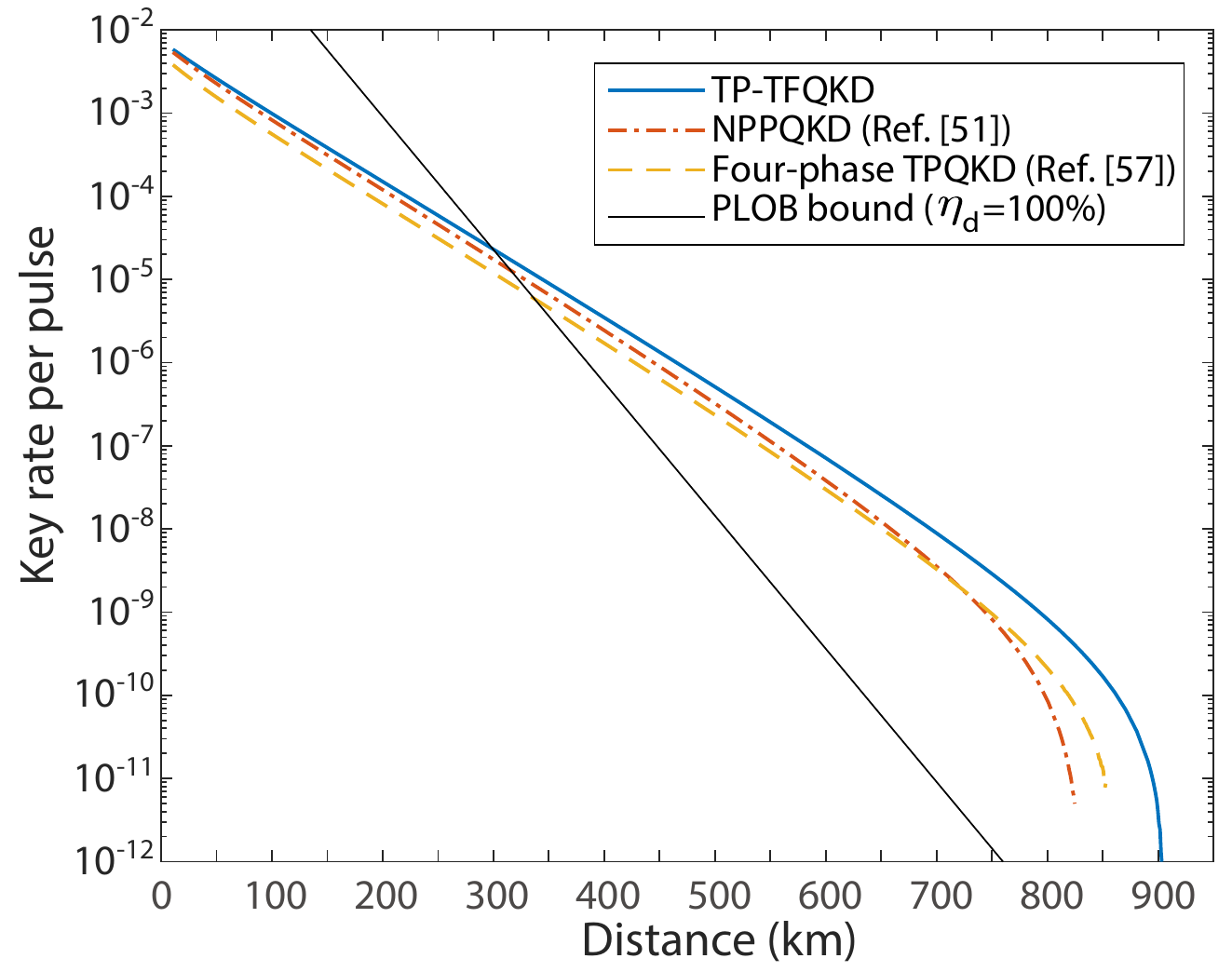}
		\caption{ Comparison of the secret key rates of TP-TFQKD, NPPQKD~\cite{maeda2019repeaterless} and four-phase TFQKD~\cite{wang2022twin} using the same parameters in the ultra-low-loss fiber scenario of Ref.~\cite{wang2022twin}. The total number of transmitted quantum signals is $3.2\times 10^{14}$. The detection efficiency is $28\%$, the dark count rate is $3\times 10^{-11}$, and the channel transmittance from Alice (Bob) to Charlie is $\eta_{a(b)} = 10^{-0.16(l-10.7)/20}$. The misalignment error rate of NPPQKD and four-phase TFQKD is $2\%$. The misalignment angle of TP-TFQKD is set as $5^{\circ}$, which contributes an error rate of approximately $2\%$. The security bounds are both $2^{-31}$. The simulations show that TP-TFQKD has higher key rates than NPPQKD and four-phase TFQKD, and that the transmission distance of TP-TFQKD can exceed 900~km.}\label{fig_tpnpp}
	\end{figure}
	
	\begin{figure}[t!]
		\centering
		\includegraphics[width=8.6cm]{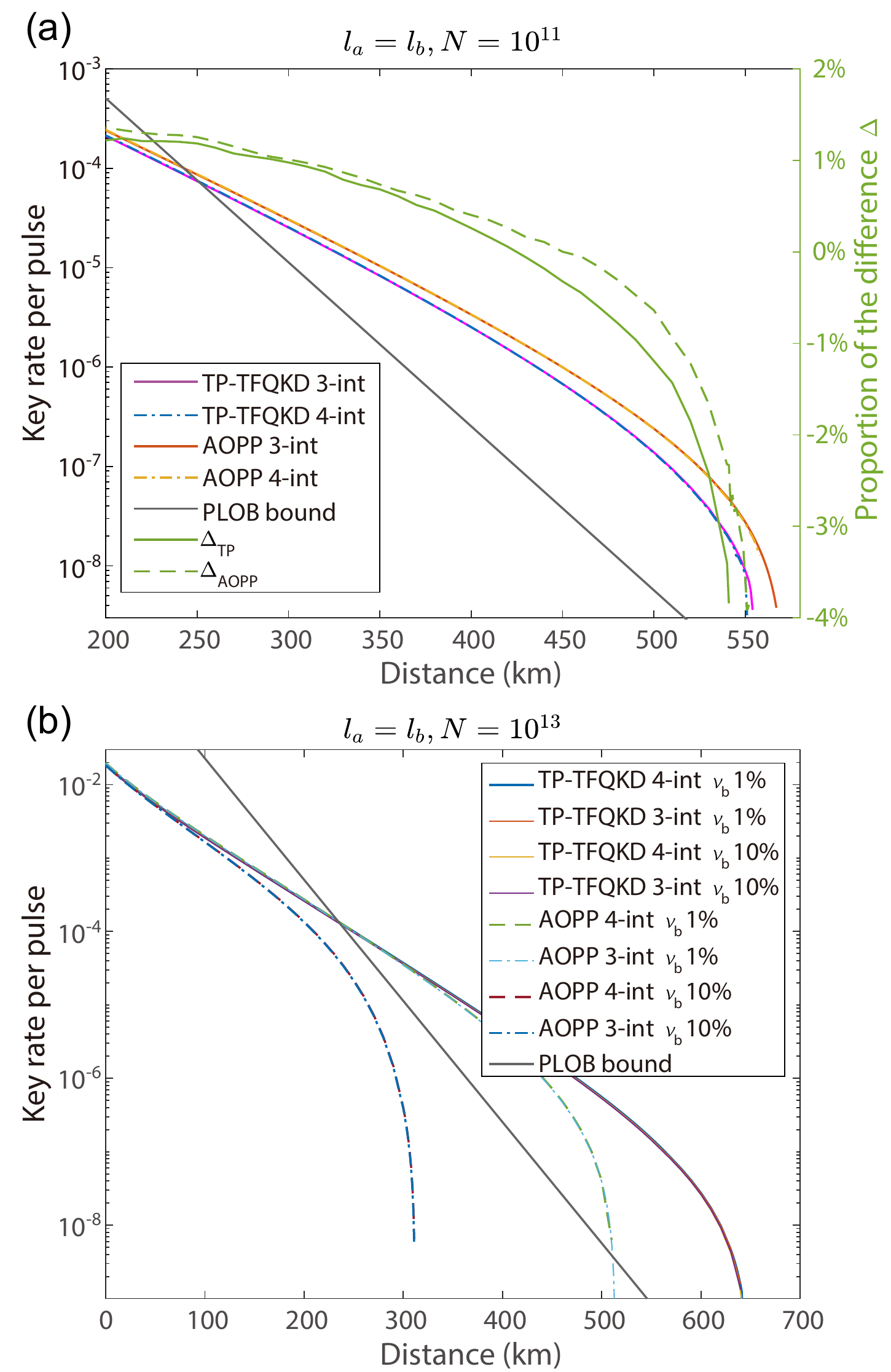}
		\caption{(a) Comparison of secret key rates between our protocol and AOPP with four-intensity and three-intensity decoy-state methods where $N = 10^{11}$. The horizontal axis represents the total transmission distance, $l=l_a+l_b$. The misalignment angle in the $X$ basis $\sigma=5^{\circ}$.
			Here, the solid (dashed) yellow line represents the relative difference between four- and three-intensity decoy states TP-TFQKD (AOPP).
			(b) Comparison of secret key rates between our protocol and AOPP with four-intensity and three-intensity decoy-state methods in the presence of a deviation in the modulated intensity, $\nu_b$, where $N = 10^{13}$.
			Our results show that the key rate of the three-intensity decoy-state method is almost the same as that of the four-intensity decoy-state method for both TP-TFQKD and AOPP, regardless of whether there is a deviation in the light intensity.}\label{fig_keyrate_fourvsthree}
	\end{figure}

In Table~\ref{tab4}, we simulated TP-TFQKD and PMQKD performance~\cite{zeng2020symmetry} when the misalignment angle in the network was $\sigma=18^{\circ}$, resulting in an interference error rate of approximately $5\%$. The results show that when there is a moderate interference error, the key rate of TP-TFQKD can still exceed the PLOB bound for a transmission distance of 400 km while PMQKD is unable to generate a secure key. In this network, TP-TFQKD enables the key rates of the two links (B-D and A-B) to exceed the PLOB bound. However, PMQKD failed to exceed the PLOB bound under the same circumstances. When the transmission distance is 300~km, the key rate of TP-TFQKD can still reach $10^{-6}$. This means that a key rate of several kbits per second can be achieved with a state-of-the-art high-speed QKD system working at a gigahertz repetition rate.

\subsection{Comparison with TFQKD protocol variants}
\label{sec_performance_APMDI}
We denote the distance between Alice (Bob) and Charlie as $l_a$~($l_b$). In Fig.~\ref{figureaoppun_fig}, we plot the secret key rates of TP-TFQKD and AOPP~\cite{jiang2020zigzag} in symmetric ($l_a=l_b)$ and asymmetric ($l_b-l_a=100$ km) channels under data size $N=10^{13}$, and we also consider the inaccuracy of the modulated light intensities. For TP-TFQKD, we use $\Delta_{\nu_b=10\%}=(R_{\rm{tp}}-R_{\delta \nu_b=10})/R_{\rm{tp}}$ to represent the relative difference between the key rate in the ideal case, $R_{\rm{tp}}$, and the key rate when the deviation of intensity $\nu_b$ is 10\% $R_{\delta \nu_b=10}$. The key rates of TP-TFQKD when the light intensity is inaccurately modulated are only slightly smaller than those when the light intensity is perfectly modulated. When the light intensity is perfectly modulated, the AOPP key rate is slightly higher than that of TP-TFQKD. However, whether in a symmetric or asymmetric channel, this advantage of AOPP no longer exists when there is a deviation of even 1\% in the light intensity preparation, $\nu_b$, and other parameters do not deviate. Figure~\ref{figureaoppun_fig}(a) shows that the transmission distance of TP-TFQKD in a symmetric channel is 200~km farther than that of AOPP when the deviations of intensity $\nu_b$ for both protocols is 3\%. Moreover, the transmission distance of the AOPP cannot exceed 320 km with a deviation of more than 10\%. Figure~\ref{figureaoppun_fig}(b) reveals that the transmission distance of TP-TFQKD is 200~km farther than that of AOPP in an asymmetric channel when the deviations of intensity $\nu_b$ for both protocols is 10\%. The key rate of AOPP is slightly better in the asymmetric case when the intensity deviations are the same because the fidelity between joint density matrices of the single-photon states in the Z and X bases is the lowest in the symmetric case for a fixed intensity deviation (see Eq.~\eqref{B_3}).

In fact, a preparation deviation of 3\% is common, especially for small values. For example, when light intensity $\nu$ prepared by user D is changed from 0.010 to 0.0103 (see Table~\ref{tab3}), the preparation deviation reaches 3\%. This preparation deviation further reduces the key rate, even resulting in a secure key rate of zero. To illustrate the importance of this point, we analyze the potential secret key in practical situations under reported experimental performance and security parameters~\cite{liu2021field,chen2021twin}.
To ensure that the AOPP key rate for the 428 km of deployed commercial fiber in Ref~\cite{liu2021field} is not zero, the preparation deviation of the $\nu_a$ light intensity is required to be less than $0.09\%$. To ensure that the AOPP key rate for the 511-km of ultra-low-loss fiber in Ref~\cite{chen2021twin} is not zero, the preparation deviation of the $\nu_a$ light intensity must be less than $0.2\%$. These two small deviations are usually difficult to achieve.

In addition, we compare the key rates of TP-TFQKD with PMQKD~\cite{zeng2020symmetry} protocols under the finite-size case in Fig.~\ref{fig_keyrate_finite_sym}. The experimental parameters are listed in Table~\ref{tab1}. The angle of misalignment in the $X$ basis of TP-TFQKD and PMQKD are both set to $\sigma=5^{\circ}$, which contributes to an error of approximately $2\%$. Figures~\ref{fig_keyrate_finite_sym}(a) and \ref{fig_keyrate_finite_sym}(b) show a comparison of key rates between TP-TFQKD and PMQKD protocols under symmetrical and asymmetrical channels, respectively. Under the same experimental parameters, TP-TFQKD can achieve significantly higher key rates and longer transmission distances than PMQKD in both symmetric and asymmetric channels. At 400 km and $N={10}^{11}$, the key rate of our protocol is 5.2 times that of PMQKD in the symmetric channel and 14 times that of PMQKD in the asymmetric channel. Compared with symmetric channels, TP-TFQKD has a greater advantage in key rates than PMQKD under asymmetric channels.

We also compare the key rates of TP-TFQKD with NPPQKD and four-phase TFQKD in the	finite-key scenario in Fig.~\ref{fig_tpnpp}. The same parameters in the ultra-low-loss fiber scenario of Ref.~\cite{wang2022twin}  are used along with data size $N=3.2\times 10^{14}$, detection efficiency $\eta_d=28\%$, dark count rate $p_d=3\times 10^{-11}$, and channel transmittance from Alice (Bob) to Charlie $\eta_{a(b)} = 10^{-0.16(l-10.7)/20}$. The misalignment error rate of NPPQKD and four-phase TFQKD $e_m=2\%$, and the misalignment angle of TP-TFQKD is set to $\sigma=5^{\circ}$, which contributes to an error rate of approximately $2\%$. The security bounds are both $2^{-31}$.
Our simulation results for the key rates of NPPQKD and four-phase TFQKD are consistent with the simulation results in Ref.~\cite{wang2022twin}. Under these system parameters, the experimental transmission distance of four-phase TFQKD achieves 833.80~km. The simulations show that if using the same system parameters, TP-TFQKD has higher key rates than NPPQKD and four-phase TFQKD, and the transmission distance of TP-TFQKD can exceed 900~km. The key rate of TP-TFQKD is approximately $10$ times that of NPPQKD when the fiber length is $800$~km. For an 830-km transmission distance, the key rate of TP-TFQKD is $3.337\times 10^{-10}$ per quantum pulse, which is approximately $11$ times that of four-phase TFQKD~($3.113\times 10^{-11}$).

\subsection{Comparison of three-intensity and four-intensity protocols}

In this section, we discuss the performance of the TP-TFQKD protocol using the decoy-state method with different intensities. The parameters used are listed in Table~\ref{tab1}. In the four-intensity decoy-state TP-TFQKD protocol, Alice chooses random phase $\theta_{a}^{i}$ $\in[0,2\pi)$ and random classical bit $r_{a}^{i}\in\{0, 1\}$ at each time bin $i\in\{1,2,\ldots,N\}$. She then prepares a weak coherent pulse $\ket{e^{\textbf{i}(\theta_{a}^{i}+r_{a}^{i}\pi)}\sqrt{k_{a}^i}}$ with probability $p_{k_a}$, where $k_a^i\in \{\mu_a, \omega_a,\nu_a, \mathbf{o}_{a}, \hat{\mathbf{o}}_{a}\}$. The intensities satisfy the relation $\mu_a>\omega_a>\nu_a>\mathbf{o}_{a}
	=\hat{\mathbf{o}}_{a} = 0$. Likewise, Bob prepares a phase-randomized weak coherent pulse: $\ket{e^{\textbf{i}(\theta_{b}^{i}+r_{b}^{i}\pi)}\sqrt{k_{b}^i}}$ ($k_b^i\in \{\mu_b,\omega_b,\nu_a, \mathbf{o}_{b}, \hat{\mathbf{o}}_{b}\}$).  The events in the $Z$ basis consist of $\{\mu_a \mathbf{o}_{a}, \mathbf{o}_{b}\mu_b\}$, $\{\mu_a\mathbf{o}_{a}, \mu_b\mathbf{o}_{b}\}$, $\{\mathbf{o}_{a}\mu_a, \mathbf{o}_{b}\mu_b\}$, and $\{\mathbf{o}_{a}\mu_a, \mu_b\mathbf{o}_{b}\}$. The events in the $X$ basis are $\{\nu_a^i\nu_a^j,\nu_b^i\nu_b^j\}$ that satisfy $\theta^{i(j)} \in[-\delta,\delta]\cup[\pi-\delta,\pi+\delta]$. At the same time, $\left|\theta^i -\theta^j\right|= 0 $ or $\pi$. Detection events  $\{o_a, \omega_b\}$, $\{\omega_a, o_b\}$, $\{\hat{\mathbf{o}}_a, \mu_b\}$, $\{\mu_a, \hat{\mathbf{o}}_b\}$, $\{\hat{\mathbf{o}}_a,o_b\}$, and $\{\mathbf{o}_a, \hat{\mathbf{o}}_b\}$ will be used later for decoy analysis. We compared the performance of the TP-TFQKD protocol using the decoy-state method with different intensities.

Figure~\ref{fig_keyrate_fourvsthree} shows comparison results of the key rates of the three- and four-intensity decoy-state protocols. In Fig.~\ref{fig_keyrate_fourvsthree}(a), the key rates between our protocol and AOPP are compared with four-intensity and three-intensity decoy-state methods under $N = 10^{11}$. $\Delta=(R^{\rm{four}}-R^{\rm{three}})/R^{\rm{four}}$ is used to represent the relative difference between the key rate of the four-intensity decoy-state protocol, $R^{\rm{four}}$, and the key rate of the three-intensity decoy-state protocol, $R^{\rm{three}}$. The results indicate that at relatively short distances, no obvious increase in key rate is observed when using more decoy intensities. Interestingly, the three-intensity protocol could provide higher key rates than the four-intensity protocol at long distances. In Fig.~\ref{fig_keyrate_fourvsthree}(b), we compare the key rates between our protocol and AOPP with four-intensity and three-intensity decoy-state methods when the intensities of the weak decoy states, $\nu_b$, of both protocols deviate under $N = 10^{13}$.
Figure~\ref{fig_keyrate_fourvsthree}(b) reveals that the gap between three-intensity decoy-state protocols and four-intensity decoy-state protocols is also small in the presence of a deviation in the modulated intensity. Hence, practical implementations of TP-TFQKD or AOPP using three intensities can be convenient and efficient.

\section{Conclusion}
\label{sec_conclusion}
In summary, we provided a two-photon (TP) TFQKD protocol through time multiplexing that can break the secret key capacity with the same experimental requirements as other TFQKD protocols. In addition, we quantified the security of the SNS-TFQKD protocol without satisfying constraints regarding intensity and probability by transforming the imperfect modulation into an information leakage problem. Compared with SNS-TFQKD (AOPP), which can theoretically transmit the farthest distance at present, our protocol achieves a longer transmission distance in the presence of deviations in modulating intensity. In a scalable network with a large number of users, our protocol compares favorably with SNS-TFQKD (AOPP) because it is unaffected by the addition or deletion of users and has an order of magnitude higher key rate than PMQKD. Furthermore, we studied the performance of TP-TFQKD and SNS-TFQKD (AOPP) using the decoy-state method with different intensities. The results show that the three-intensity protocol of both TP-TFQKD and SNS-TFQKD (AOPP) achieve a good trade-off between technical complexity and performance.

In contrast to the phase-matching-type QKD protocol, TP-TFQKD can further improve the key rate by harnessing single-photon sources or other nonclassical light sources because the non-interference mode is used for key generation. Our protocol has important practical advantages: high misalignment error tolerance, long transmission distance, and independent setting of probability and intensity for each user. Thus, our protocol is highly versatile and suitable for most metropolitan networks. The post-matched two-photon interference performed in our scheme can also apply to other quantum information processing, such as quantum conference key agreement, quantum secret sharing, and quantum digital signatures. 

Note added--During  the time our work was posted on the preprint server,  we notice that  Ref.~\cite{Jiang2022Robust} also solve the problem for the SNS-TFQKD protocol where  source errors are unavoidable.  Additionally, two different research studies exploiting post-matching ideas to make MDIQKD break the repeaterless rate-transmittance bound have been reported~\cite{xie2022breaking,zeng2022mode}. In this work, we propose a post-matching TF-type protocol tailored for scalable TFQKD networks and show that it is secure against coherent attacks in the finite-key regime. Although our post-matching ideas are similar to these recently published works, there are differences in the density matrix of single-photon pairs and security in the protocols.

\section*{Acknowledgement}
We gratefully acknowledge the support from the National Natural Science Foundation of China (No. 12274223), the Natural Science Foundation of Jiangsu Province (No. BK20211145), the Fundamental Research Funds for the Central Universities (No. 020414380182), the Key Research and Development Program of Nanjing Jiangbei New Area (No. ZDYD20210101),  the Program for Innovative Talents and Entrepreneurs in Jiangsu (JSSCRC2021484), the Program of Song Shan Laboratory (Included in the management of Major  Science and Technology Program of Henan Province) (No. 221100210800-02), and the China Postdoctoral Science Foundation (No. 2021M691536).

\onecolumngrid
\appendix

\section{Security analysis for TP-TFQKD protocol}\label{secur_TP}
For our practical protocol, the normalized density matrix of the joint quantum state with a single-photon pair in the $Z$ basis is
\begin{equation}
	\begin{aligned}
		\boldsymbol{\rho}_{z}^{11}&=\frac{1}{2}\left(\ket{+z-z}_{ab}\bra{+z-z}+\ket{-z+z}_{ab}\bra{-z+z}\right)\\
		&=\frac{1}{2}\left(\ket{10}_{a}^{ij}\bra{10}\otimes\ket{01}_{b}^{ij}\bra{01}+\ket{01}_{a}^{ij}\bra{01}\otimes\ket{10}_{b}^{ij}\bra{10}\right)\\
		&=\frac{1}{2}\left(\ket{10}_{ab}^{i}\bra{10}\otimes\ket{01}_{ab}^{j}\bra{01}+\ket{01}_{ab}^{i}\bra{01}\otimes\ket{10}_{ab}^{j}\bra{10}\right).\\
	\end{aligned}
\end{equation}
This requires Alice to prepare $\ket{0}$ at one time bin and $\ket{1}$ at the other time bin, whereas Bob does the opposite. We consider this ``effective,'' which will be used later. Note that the measurement outcomes of quantum states $\ket{10}_{a}^{ij}\ket{10}_{b}^{ij}$ and $\ket{01}_{a}^{ij}\ket{01}_{b}^{ij}$ are not ``effective'' events.

After matching time bin $i$, where the global phases of Alice and Bob are both $\kappa$, with time bin $j$, where the global phases of Alice and Bob are both $\varphi$, the four-mode coherent state in the $X$ basis can be written as $\ket{\Psi}=\ket{e^{i(\kappa+r_{a}^{i}\pi)}\alpha_{a}}_{a}^{i}\otimes\ket{e^{i(\varphi+r_{a}^{j}\pi)}\alpha_{a}}_{a}^{j}\otimes\ket{e^{i(\kappa+r_{b}^{i}\pi)}\alpha_{b}}_{b}^{i}\otimes\ket{e^{i(\varphi+r_{b}^{j}\pi)}\alpha_{b}}_{b}^{j}$. The security will not be affected when the global phases of Alice and Bob do not satisfy the specific correlation because the global phase noise difference is determined by an untrusted Charlie. To reduce the interference error, it is preferable to satisfy the phase correlation. 
Given that the values of $\kappa$ and $\varphi$ are both randomized, the density matrix of the four-mode joint quantum state in the $X$ basis can be written as
\begin{equation}
	\begin{aligned}\label{Aeq6}
		\hat{\rho}_{r_{a}^{i}r_{b}^{i}r_{a}^{j}r_{b}^{j}}&=\frac{1}{(2\pi)^{2}}\int_{0}^{2\pi}\int_{0}^{2\pi}\ket{\Psi} \bra{\Psi}d\kappa d\varphi\\
		&=e^{-2(\nu_a+\nu_b)}\left(\sum_{n=0}^{\infty}\sum_{m=0}^{\infty}\frac{(\nu_a+\nu_b)^{n+m}}{n!m!}\ket{n}_{ab}^{i}\bra{n}\otimes\ket{m}_{ab}^{j}\bra{m}\right),
	\end{aligned}
\end{equation}
where $\nu_a=|\alpha_a|^{2}$, $\nu_b=|\alpha_b|^{2}$ and Fock states
\begin{equation}
	\begin{aligned}
		\ket{n}_{ab}^{i}&=\frac{\left(\sqrt{\nu_a}a_{i}^{\dagger}+e^{i(r_{b}^{i}-r_{a}^{i})\pi}\sqrt{\nu_b}b_{i}^{\dagger}\right)^{n}}{\sqrt{(\nu_a+\nu_b)^{n}n!}}\ket{00}_{ab}^{i},\\
		\ket{m}_{ab}^{j}&=\frac{\left(\sqrt{\nu_a}a_{j}^{\dagger}+e^{i(r_{b}^{j}-r_{a}^{j})\pi}\sqrt{\nu_b}b_{j}^{\dagger}\right)^{m}}{\sqrt{(\nu_a+\nu_b)^{m}m!}}\ket{00}_{ab}^{j}.
	\end{aligned}
\end{equation}
The corresponding single-photon pair component is
\begin{equation}
	\begin{aligned}
		\hat{\rho}_{r_{a}^{i}r_{b}^{i}r_{a}^{j}r_{b}^{j}}^{11}&=\ket{1}_{ab}^{i}\bra{1}\otimes\ket{1}_{ab}^{j}\bra{1}\\
		&=\frac{\sqrt{\nu_a}\ket{10}_{ab}^{i}+e^{i(r_{b}^{i}-r_{a}^{i})\pi}\sqrt{\nu_b}\ket{01}_{ab}^{i}}{\sqrt{\nu_a+\nu_b}}\frac{\sqrt{\nu_a}\bra{10}_{ab}^{i}+e^{-i(r_{b}^{i}-r_{a}^{i})\pi}\sqrt{\nu_b}\bra{01}_{ab}^{i}}{\sqrt{\nu_a+\nu_b}}\\
		&\otimes\frac{\sqrt{\nu_a}\ket{10}_{ab}^{j}+e^{i(r_{b}^{j}-r_{a}^{j})\pi}\sqrt{\nu_b}\ket{01}_{ab}^{j}}{\sqrt{\nu_a+\nu_b}}\frac{\sqrt{\nu_a}\bra{10}_{ab}^{j}+e^{-i(r_{b}^{j}-r_{a}^{j})\pi}\sqrt{\nu_b}\bra{01}_{ab}^{j}}{\sqrt{\nu_a+\nu_b}}.
	\end{aligned}
\end{equation}
Hence, the density matrix of the joint quantum state with a single-photon pair in the $X$ basis is
\begin{equation}
	\begin{aligned}
		\hat{\rho}^{11}_x=&\frac{1}{16}\big(\rho_{0000}^{11}+\rho_{0011}^{11}+\rho_{0101}^{11}+\rho_{0110}^{11}+\rho_{1001}^{11}+\rho_{1010}^{11}+\rho_{1100}^{11}+\rho_{1111}^{11}\\
		&\qquad+\rho_{0001}^{11}+\rho_{0010}^{11}+\rho_{0100}^{11}+\rho_{0111}^{11}+\rho_{1000}^{11}+\rho_{1011}^{11}+\rho_{1101}^{11}+\rho_{1110}^{11}\big)\\
		=&\frac{2\nu_a\nu_b}{(\nu_a+\nu_b)^2}\frac{(\ket{\psi^{+}}_{ab}\bra{\psi^+}+\ket{\psi^{-}}_{ab}\bra{\psi^-})}{2}+\frac{\nu_a^2+\nu_b^2}{(\nu_a+\nu_b)^2}\frac{(\ket{\chi^{+}}_{ab}\bra{\chi^+}+\ket{\chi^{-}}_{ab}\bra{\chi^-})}{2},\\
	\end{aligned}
\end{equation}
where $\ket{\psi^{\pm}}_{ab}=(\ket{10}_{ab}^{i}\ket{01}_{ab}^{j}\pm\ket{01}_{ab}^{i}\ket{10}_{ab}^{j})/\sqrt{2}$ and $\ket{\chi^{\pm}}_{ab}=(\nu_a\ket{10}_{ab}^{i}\ket{10}_{ab}^{j}\pm\nu_b\ket{01}_{ab}^{i}\ket{01}_{ab}^{j})/\sqrt{\nu_a^2+\nu_b^2}$. 
We define $\boldsymbol{\rho}_{x}^{11}:=(\ket{\psi^{+}}_{ab}\bra{\psi^+}+\ket{\psi^{-}}_{ab}\bra{\psi^-})/2$ and $\tilde{\boldsymbol{\rho}}_{x}^{11}:=(\ket{\chi^{+}}_{ab}\bra{\chi^+}+\ket{\chi^{-}}_{ab}\bra{\chi^-})/2$ as density matrices of ``effective'' and ``ineffective'' single-photon pairs in the X basis, respectively.
The single-photon pair component in the $X$ basis can be regarded as a classical mixture of ``effective'' single-photon pairs and ``ineffective'' single-photon pairs:
\begin{equation}
	\begin{aligned}\label{Aeq7}
		\hat{\rho}^{11}_x
		=&\frac{2\nu_a\nu_b}{(\nu_a+\nu_b)^2}\boldsymbol{\rho}_{x}^{11}
		+\frac{\nu_a^2+\nu_b^2}{(\nu_a+\nu_b)^2}\tilde{\boldsymbol{\rho}}_{x}^{11}.
	\end{aligned}
\end{equation}
Note that the ``effective'' single-photon pairs in the $X$ and $Z$ bases have the same density matrices:
\begin{equation}
	\begin{aligned}\label{Aeq8}
		\boldsymbol{\rho}_{x}^{11}=\boldsymbol{\rho}_{z}^{11}&=\frac{1}{2}\left(\ket{\psi^{+}}_{ab}\bra{\psi^+}+\ket{\psi^{-}}_{ab}\bra{\psi^-}\right)\\
		&=\frac{1}{2}\left(\ket{10}_{ab}^{i}\bra{10}\otimes\ket{01}_{ab}^{j}\bra{01}+\ket{01}_{ab}^{i}\bra{01}\otimes\ket{10}_{ab}^{j}\bra{10}\right).\\	
	\end{aligned}
\end{equation}
Therefore, the bit error rate of ``effective'' single-photon pairs in the $X$ basis, ${e}_{11}^x$, is asymptotically equal to the phase error rate of ``effective'' single-photon pairs in the $Z$ basis, $\phi_{11}^z$. 
Combining Eqs.~\eqref{Aeq6} and \eqref{Aeq7}, we obtain that the four-mode joint quantum state sent by Alice and Bob collapses to``effective'' single-photon pairs with probability $2\nu_a\nu_be^{-2(\nu_a+\nu_b)}$; the vacuum state at the previous time bin (this case is denoted as $\mathcal{V}_{i}$) with probability $e^{-(\nu_a+\nu_b)}$; the vacuum state at the latter time bin (denoted as $\mathcal{V}_{j}$) with probability $e^{-(\nu_a+\nu_b)}$; and the vacuum states at both time bins (denoted as $\mathcal{V}_{ij}$) with probability $e^{-2(\nu_a+\nu_b)}$. The error in the $X$ basis comes from four parts: the vacuum state, the ``effective'' single-photon pair, the ``ineffective'' single-photon pair, and the multi-photon pair.
Denoting the bit error rate in the $X$ basis as $E^x$ and using the fact that the error rate of the vacuum state is always $1/2$, we can estimate the upper bound for the error rate of ``effective'' single-photon pairs in the $X$ basis by
\begin{equation}
	\begin{aligned}
		e_{11}^x\le&\frac{E^x-1/2e^{-(\nu_a+\nu_b)}(q_{0\nu_a,0\nu_b}+q_{\nu_a0,\nu_b0})+1/2e^{-2(\nu_a+\nu_b)}q_{00,00}}{2\nu_a\nu_be^{-2(\nu_a+\nu_b)}Y_{11}^x},\\
	\end{aligned}
\end{equation}
where $Y_{11}^x$ is the yield of ``effective'' single-photon pairs in the $X$ basis, $q_{0\nu_a,0\nu_b}~(q_{\nu_a0,\nu_b0})$ is the yield of $\mathcal{V}_{i}~(\mathcal{V}_{j})$, and $q_{00,00}$ is the yield of $\mathcal{V}_{ij}$. Note that $\mathcal{V}_{ij}$ is the subset of $\mathcal{V}_{i}$ and $\mathcal{V}_{j}$. The error rate caused by $\mathcal{V}_{ij}$ is subtracted twice in the term $1/2e^{-(\nu_a+\nu_b)}(q_{0\nu_a,0\nu_b}+q_{\nu_a0,\nu_b0})$, for which the term $1/2e^{-2(\nu_a+\nu_b)}q_{00,00}$ is added to the numerator as compensation. The detailed calculation of the bit error rate in the $X$ basis and the phase error rate in the $Z$ basis using the decoy-state method is given in Appendix~\ref{simu_p1}.

As shown by Eqs.~\eqref{Aeq7}–\eqref{Aeq8}, similar to MDIQKD, the density matrices of the ``effective'' single-photon pairs are always identical in the $X$ and $Z$ bases regardless of the source parameters independently chosen by Alice and Bob. Hence, virtual protocol of entanglement swapping and purification always holds. On the other hand, different from MDIQKD, the bit error rate of ``effective'' single-photon pairs in the $Z$ basis of our protocol is strictly zero. 
\twocolumngrid
 
\section{Security analysis for SNS-TFQKD protocol}\label{app_sns}
Under the SNS-TFQKD protocol~\cite{wang2018twin,hu2019sending}, when Alice~(Bob) chooses the $Z$-window, she~(he) prepares and sends the phase-randomized coherent state $\ket{\sqrt{\mu_a}}~(\ket{\sqrt{\mu_b}})$ with probability $t_a~(t_b)$ or sends a vacuum pulse with probability $1-t_a~(1-t_b)$. The corresponding density matrix of the joint single-photon states in the $Z$ basis is

\begin{equation}
	\begin{aligned}
		\rho^1_z=&C[t_a(1-t_b)\mu_ae^{-\mu_a}\ket{10}_{ab}\bra{10}\\
		+&t_b(1-t_a)\mu_be^{-\mu_b}\ket{01}_{ab}\bra{01}],
	\end{aligned}
\end{equation}
where $C=1/[t_a(1-t_b)\mu_ae^{-\mu_a}+t_b(1-t_a)\mu_be^{-\mu_b}]$ is the normalization factor. When Alice~(Bob) chooses the $X$ window, she (he) prepares and sends phase-randomized coherent state $\ket{\sqrt{\nu_a}}~(\ket{\sqrt{\nu_b}})$. After implementing post-selected phase matching, the density matrix of the joint one-photon states in the $X$ basis is
\begin{equation}
	\begin{aligned}
		\rho^1_x=&\frac{1}{\nu_a+\nu_b}[\nu_a\ket{10}_{ab}\bra{10}
		+\nu_b\ket{01}_{ab}\bra{01}].\\
	\end{aligned}
\end{equation}
As a result, one requests the following mathematical constraint for source parameters~\cite{hu2019sending}
\begin{equation}\label{mathconstra_sns}
	\begin{aligned}
		\frac{\nu_a}{\nu_b}=&\frac{t_a(1-t_b)\mu_ae^{-\mu_a}}{t_b(1-t_a)\mu_be^{-\mu_b}}\\
	\end{aligned}
\end{equation}
to satisfy the relation $\rho^1=\rho^1_x=\rho^1_z$ so that one can use the bit error rate in the $X$ basis to estimate the phase error rate in the $Z$ basis. Note that one always needs to satisfy Eq.~\eqref{mathconstra_sns}, even for symmetric channels ~\cite{wang2018twin}. 

However, Eq.~\eqref{mathconstra_sns} is not usually satisfied in a multiuser quantum network. In this case, to provide a secure key rate, we introduce the imbalance of the quantum coin $\Delta$, defined as~\cite{lo2007security,koashi2009simple}
\begin{equation}
	\begin{aligned}\label{B_3}
		\Delta&=\frac{1}{2Q_{1}}[1-F(\rho_{x}^1,\rho_{z}^1)],\\
	\end{aligned}
\end{equation}
where $Q_1=C[t_a(1$$-$$t_b)\mu_ae^{-\mu_a} y_{10}+t_b(1-t_a)\mu_be^{-\mu_b}y_{01}]$ is the yield of single photons in the $Z$ basis and $\rho_{x}^{1}$ ($\rho_{z}^{1}$) is the joint density matrix of single-photon states in the $X$ ($Z$) basis prepared by Alice and Bob. $F(\rho_{x},\rho_{z})={\rm tr}\sqrt{\rho_{x}^{1/2}\rho_{z}\rho_{x}^{1/2}}$ is the fidelity between density matrices $\rho_{x}$ and $\rho_{z}$. With the relationship $1-2\Delta\leq\sqrt{(1-e_{1}^{x*})(1-e_{\rm ph}^{z*})}+\sqrt{e_{1}^{x*}e_{\rm ph}^{z*}}$, the upper bound of the expected value of the phase error rate in the $Z$ basis $e_{\rm ph}^{z*}$~\cite{lo2007security} can be expressed as
\begin{equation}
	\begin{aligned}\label{B_4}
		\overline{e}_{\rm ph}^{z*}\le &\left[(1-2\Delta)\sqrt{e_{1}^{x*}}+2\sqrt{\Delta(1-\Delta)(1-e_{1}^{x*})}\right]^2\\
		\le& e_{1}^{x*}+4{\Delta}+4\sqrt{{\Delta} {e}_{1}^{x*}},
	\end{aligned}
\end{equation}
where $e_{1}^{x*}$ is the expected value of the bit error rate in the $X$ basis.
From Eq.~\eqref{B_3} and Eq.~\eqref{B_4}, we can see that $\Delta$ increases dramatically as distance increases, leading to extremely poor estimation of the phase error rate $\overline{e}_{\rm ph}^{z}$.

\section{Detailed calculation of TP-TFQKD protocol}\label{app_simulation}
\subsection{Calculation methods }\label{simu_p1}
In this section, we calculate the parameters in Eq. \eqref{eq_keyrate_finite} to estimate the secret key rate in the finite-size regime. We apply the variant of the Chernoff bound to obtain the upper and lower bounds of expected value $x^{*}$ for a given observed value $x$ and failure probability $\varepsilon$~\cite{yin2020tight}:
\begin{equation}
	\begin{aligned}\label{varchernoff1}
		\overline{x}^{*}&=x+\beta+\sqrt{2\beta x+\beta^{2}}\\
		\underline{x}^{*}&=\max\left\{x-\frac{\beta}{2}-\sqrt{2\beta x+\frac{\beta^{2}}{4}},~0\right\},\\
	\end{aligned}
\end{equation}
where $\beta=\ln{\varepsilon^{-1}}$. We use the Chernoff bound to obtain the upper and lower bounds of observed value $x$ for a given expected value $x^{*}$ and failure probability $\varepsilon$~\cite{yin2020tight}:
\begin{equation}
	\begin{aligned}\label{chernoff1}
		\overline{x}&=x^{*}+\frac{\beta}{2}+\sqrt{2\beta x^{*}+\frac{\beta^{2}}{4}}\\
		\underline{x}&=x^{*}-\sqrt{2\beta x^{*}}.
	\end{aligned}
\end{equation} 
We denote the number of $\{k_a,~k_b\}$ as $x_{k_ak_b}$. We denote the number and error number of events $\{k_a^{i}k_a^{j},~k_b^{i}k_b^{j}\}$ after post-matching as $n_{k_a^{i}k_a^{j},~k_b^{i}k_b^{j}}$ and $m_{k_a^{i}k_a^{j},~k_b^{i}k_b^{j}}$, respectively. For simplicity, we abbreviate $k_a^ik_a^j,k_a^ik_a^j$ as $2k_a,2k_b$ when $k_a^i=k_a^j$ and $k_b^i=k_b^j$.

(1)~$\underline{s}_{11}^{z}$.
	$s_{11}^z$ corresponds to the number of successful detection events, where Alice and Bob emit a single photon in different time bins in the $Z$ basis. Define $y_{10}$ ($y_{01}$) as the yield of events in which Alice (Bob) emits a single photon and Bob (Alice) emits a vacuum state in an $\{\mu_a,\mathbf{o}_b\}$ ($\{\mathbf{o}_a,\mu_b\}$) event. The lower bounds of their expected values can be estimated using the decoy-state method:
	\begin{align}
		\underline{y}_{01}^*\geq& \frac{\mu_b}{N(\mu_b\nu_b-\nu_b^2)}\times\\ 
  \nonumber &\left(\frac{e^{\nu_b}\underline{x}_{o_a\nu_b}^{*}}{p_{o_a}p_{\nu_b}} -\frac{\nu_b^2}{\mu_b^2}  \frac{e^{\mu_b}\overline{x}_{\hat{\mathbf{o}}_{a}\mu_b}^{*}}{p_{\hat{\mathbf{o}}_{a}}p_{\mu_b}} - \frac{\mu_b^2-\nu_b^2}{\mu_b^2}
 \frac{\overline{x}_{o o}^{d*}}{p_{oo}^d}\right),\\
    	\underline{y}_{10}^*\geq&\frac{\mu_a}{N(\mu_a\nu_a-\nu_a^2)} \times\\ \nonumber
     &\left(  \frac{e^{\nu_a}\underline{x}_{\nu_ao_b}^{*}}{p_{\nu_a}p_{o_{b}}}-\frac{\nu_a^2}{\mu_a^2} \frac{e^{\mu_a}\overline{x}_{\mu_a\hat{\mathbf{o}}_b}^{*}}{p_{\mu_a}p_{\hat{\mathbf{o}}_{b}}}- \frac{\mu_a^2-\nu_a^2}{\mu_a^2}\frac{\overline{x}_{o o}^{d*}}{p_{oo}^d}\right),\label{eq_decoy_Y01}
	\end{align}
	where $x_{oo}^{d}=x_{\hat{\mathbf{o}}_{a}\hat{\mathbf{o}}_{b}}+x_{\hat{\mathbf{o}}_{a}\mathbf{o}_b}+x_{\mathbf{o}_{a}\hat{\mathbf{o}}_{b}}$ represents the number of events where at least one user chooses the declare-vacuum state and $p_{oo}^d=p_{\hat{\mathbf{o}}_{a}}p_{\hat{\mathbf{o}}_{b}}+p_{\hat{\mathbf{o}}_{a}}p_{\mathbf{o}_b}+p_{\mathbf{o}_{a}}p_{\hat{\mathbf{o}}_{b}}$ refers to the corresponding probability. Let $z_{10}$   ($z_{01}$)  denotes the number of  events in which Alice (Bob) emits a single photon and Bob (Alice) emits a vacuum state in an $\{\mu_a,\mathbf{o}_b\}$ ($\{\mathbf{o}_a,\mu_b\}$) event. The lower bounds of their expected values are $\underline{z}_{10}^{*}= N p_{\mu_a} p_{\mathbf{o}_{b}} \mu_{a}e^{-\mu_a} \underline{y^*_{10}}$ and $
\underline{z}^*_{01} = N p_{\mathbf{o}_{a}} p_{\mu_b} \mu_{b} e^{-\mu_b} \underline{y^*_{01}} $, respectively. Thus, the expected lower bound of the ``effective'' single-photon pairs in the $Z$ basis is given by
\begin{equation}
	\begin{aligned}
		\underline{s}_{11}^{z*}&=x_{\rm min
        } \frac{\underline{z}_{01}^*}{x_{\mathbf{o}_a\mathbf{o}_b}+x_{\mathbf{o}_a\mu_b}}\frac{\underline{z}_{10}^*}{x_{\mu_a\mathbf{o}_b}+x_{\mu_a\mu_b}}\\
        &=\frac{\underline{z}_{01}^*\underline{z}_{10}^*}{x_{\rm max}}
	\end{aligned}
\end{equation}
where $x_{\min}=\min\{x_{\mathbf{o}_a\mathbf{o}_b}+x_{\mathbf{o}_a\mu_b},x_{\mu_a\mathbf{o}_b}+x_{\mu_a\mu_b}\}$ and $x_{\max}=\max\{x_{\mathbf{o}_a\mathbf{o}_b}+x_{\mathbf{o}_a\mu_b},x_{\mu_a\mathbf{o}_b}+x_{\mu_a\mu_b}\}$.

(2)~$\underline{s}_{0\mu_b}^{z}$. $s_{0\mu_b}^z$ represents the number of events in the $Z$ basis when Alice emits a zero-photon state in the two matched time bins and the total intensity of Bob's pulses is $\mu_b$. Let  $z_{00}$ ($z_{0\mu_b}$) represents the number of detection events   where the state sent by Alice collapses to the vacuum state in the $\{\mu_a,\mathbf{o}_b\}$ ($\{\mu_a,\mu_b\}$) event. The lower bound of the expected values is $\underline{z}_{00}^*={ p_{\mu_a} p_{\mathbf{o}_b}e^{-\mu_a}\underline{x}_{o o}^{d*}}/{p_{o_ao_b}^d}$ and $ ~\underline{z}_{0\mu_b}^*=  {p_{\mu_a} p_{\mu_b}e^{-\mu_a}\underline{x}_{\mathbf{o}_{a}\mu_b}^{*}}/{p_{\mathbf{o}_{a}}p_{\mu_b}}$, respectively. Here, we utilize the relationship between the expected value $\underline{x}_{\mathbf{o}_{a}\mu_b}^{*}={ p_{\mathbf{o}_{a}} \underline{x}_{\hat{\mathbf{o}}_{a}\mu_b}^{*}}/{ p_{\hat{\mathbf{o}}_{a}}}$, and $~\underline{x}_{\mathbf{o}_{a}\mathbf{o}_b}^{*}={ p_{\mathbf{o}_{a}} p_{\mathbf{o}_b}\underline{x}_{o o}^{d*}}/{p_{oo}^d}$.   The lower bound of $s_{0\mu_b}^{z*}$ can be written as
\begin{equation}
	\begin{aligned}
		\underline{s}_{0\mu_b}^{z*}=& x_{\rm min
        } \frac{\underline{x}_{\mathbf{o}_a\mu_b}^*}{x_{\mathbf{o}_a\mathbf{o}_b}+x_{\mathbf{o}_a\mu_b}}\frac{\underline{z}_{00}^*}{x_{\mu_a\mathbf{o}_b}+x_{\mu_a\mu_b}}\\
        + &x_{\rm min }\frac{x_{\mathbf{o}_a\mathbf{o}_b}^*}{\underline{x}_{\mathbf{o}_a\mathbf{o}_b}+x_{\mathbf{o}_a\mu_b}}\frac{\underline{z}_{0\mu_b}^*}{x_{\mu_a\mathbf{o}_b}+x_{\mu_a\mu_b}}\\
=&\frac{\underline{x}_{\mathbf{o}_a\mu_b}^*\underline{z}_{00}^*+x_{\mathbf{o}_a\mathbf{o}_b}^*\underline{z}_{0\mu_b}^*}{x_{\rm max}}.
	\end{aligned}
\end{equation}

(3)~$\underline{s}_{11}^{x}$.
We define the global phase difference between Alice and Bob as $\theta= \theta_a- \theta_b +\phi_{ab}$. All events in the $X$ basis can be grouped according to phase difference $\theta~(\in\{-\delta,\delta\}\cup\{\pi-\delta,\pi+\delta\})$. In the post-matching step, two time bins are matched if they have the same global phase difference, $\theta$. Suppose $\theta$ is a randomly and uniformly distributed value and consider the angle of misalignment in the $X$ basis, $\sigma$, the expected lower bound of the ``effective'' single-photon pairs in the $X$ basis can then be given by
\begin{equation}
	\begin{aligned}
		\underline{s}_{11}^{x*} &=4\int_\sigma^{\sigma+\delta} \frac{N  p_{\nu_a}p_{\nu_b}q_{\nu_a\nu_b}^{\theta}}{4\pi} \times  \frac{2\nu_a\nu_be^{-2(\nu_a+\nu_b)}\underline{Y}_{11}^{x*}}{(q_{\nu_a\nu_b}^\theta)^2}d \theta\\
		&=\frac{2Np_{\nu_{a}}p_{\nu_{b}}\nu_a\nu_be^{-2(\nu_a+\nu_b)}\underline{y}_{01}^*\underline{y}_{10}^*}{\pi}\int_\sigma^{\sigma+\delta}\frac{1}{q_{\nu_a\nu_b}^\theta}d\theta,\\
	\end{aligned}
\end{equation}
where $q_{\nu_a \nu_b }^\theta$ is the gain when Alice chooses $\nu_a $ and Bob chooses $\nu_b$ with phase difference $\theta$. In the second equation, we use the relationship implied in Eq.~\eqref{Aeq8}: $Y_{11}^{x*}=Y_{11}^{z*}=y_{01}^*y_{10}^*$, where $Y_{11}^{z*}$ is the expected yield of the ``effective'' single-photon pairs in the $Z$ basis.

(4)~$\overline{e}_{11}^{x}$. 
For ``effective'' single-photon pairs, the expected value of the phase error rate in the $Z$ basis is equal to the expected value of the bit error rate in the $X$ basis. Therefore, we first calculate the number of errors of the ``effective'' single-photon pairs in the $X$ basis ${t_{11}^x}$. The upper bound of ${t_{11}^x}$ can be expressed as 
\begin{equation}
	\begin{aligned}
		\overline{t}_{11}^x\leq& m^x - \underline{(m_{\nu_a0,\nu_b0}+m_{0\nu_a,0\nu_b})} +\overline{m}_{00,00}, 	
	\end{aligned}\label{eq_TPTFQKD _t11}
\end{equation}
where $m_{\nu_a0,\nu_b0}$ ($m_{0\nu_a,0\nu_b}$) is the error count when the states sent by Alice and Bob in time bin $i$~($j$) both collapse to the vacuum state in events $\{2\nu_a,2\nu_b\}$, and $m_{00,00}$ corresponds to the event where the states sent by Alice and Bob both collapse to vacuum states in events $\{2\nu_a,2\nu_b\}$. The expected counts $\underline{(n_{\nu_a0,\nu_b0}+n_{0\nu_a,0\nu_b})}^*$ and $\overline{n}_{00, 00}^*$ can be expressed as
\begin{equation}
	\begin{aligned}			&\underline{(n_{\nu_a0,\nu_b0}+n_{0\nu_a,0\nu_b})}^*\\
 =&4\int_\sigma^{\sigma+\delta} \frac{N  p_{\nu_a}p_{\nu_b}q_{\nu_a\nu_b}^{\theta}}{4\pi} \times \frac{e^{-(\nu_a+\nu_b)}(\underline{q}_{0\nu_a,0\nu_b}^{\theta*}+\underline{q}_{\nu_a0,\nu_b0}^{\theta*})}{(q_{\nu_a\nu_b}^\theta)^2}d \theta\\
		=&\frac{2\delta Np_{\nu_{a}}p_{\nu_{b}}e^{-(\nu_a+\nu_b)}\underline{q}_{00}^{*}}{\pi}\\
	\end{aligned}
\end{equation}
and 
\begin{equation}
	\begin{aligned}	
		\overline{n}_{00, 00}^*=&4\int_\sigma^{\sigma+\delta} \frac{N  p_{\nu_a}p_{\nu_b}q_{\nu_a\nu_b}^{\theta}}{4\pi} \times \frac{e^{-2(\nu_a+\nu_b)}\overline{q}_{00,00}^*}{(q_{\nu_a\nu_b}^\theta)^2}d \theta\\
		=&\frac{Np_{\nu_{a}}p_{\nu_{b}}e^{-2(\nu_a+\nu_b)}(\overline{q}_{00}^{*})^2}{\pi}\int_\sigma^{\sigma+\delta}\frac{1}{q_{\nu_a\nu_b}^\theta}d\theta\\
	\end{aligned}
\end{equation}
respectively, where we use the relationships $q_{0\nu_a,0\nu_b}^{\theta*}=q_{\nu_a0,\nu_b0}^{\theta*}=q_{00}^*q_{\nu_a\nu_b}^\theta$, $q_{00,00}^*=(q_{00}^*)^2$, and $q_{00}^*=x_{o_ao_b}^{d*}/(Np_{oo}^d)$. Then, we have $\underline{(m_{\nu_a0,\nu_b0}+m_{0\nu_a,0\nu_b})}^*=\frac{1}{2}\underline{(n_{\nu_a0,\nu_b0}+n_{0\nu_a,0\nu_b})}^*$ and $\overline{m}_{00, 00}^*=\frac{1}{2}\overline{n}_{00, 00}^*$ since the error rate of the vacuum state is always $1/2$. The upper bound of the bit error rate in the $X$ basis can be given by $\overline{e}_{11}^{x}=\overline{t_{11}^{x}}/\underline{s}_{11}^x$.	

(5)~$\overline{\phi}_{11}^{z}$.
Finally, for failure probability $\epsilon$, the upper bound of the phase error rate, $\phi_{11}^{z}$, can be obtained using random sampling without replacement~\cite{yin2020tight}:
\begin{equation}
	\begin{aligned}
		\overline{\phi}_{11}^{z}\leq&\overline{e}_{11}^{x}+	\gamma^U \left(\underline{s}_{11}^z,\underline{s}_{11}^x,\overline{e}_{11}^{x},\epsilon\right),\\
	\end{aligned}\label{eq_TPTFQKD _phi11z}
\end{equation}
where
\begin{equation}
	\gamma^{U}(n,k,\lambda,\epsilon)=\frac{\frac{(1-2\lambda)AG}{n+k}+
		\sqrt{\frac{A^2G^2}{(n+k)^2}+4\lambda(1-\lambda)G}}{2+2\frac{A^2G}{(n+k)^2}},
\end{equation}
with $A=\max\{n,k\}$ and $G=\frac{n+k}{nk}\ln{\frac{n+k}{2\pi nk\lambda(1-\lambda)\epsilon^{2}}}$. 

For calculation of the asymptotic key rate, we just need to let the expected value be equal to the observed value and let the phase error rate in the $Z$ basis be equal to the bit error rate in the $X$ basis.

\subsection{Simulation formulas}\label{simu_pb}
When Alice and Bob send intensities $k_a$ and $k_b$ with phase difference $\theta$, the gain corresponding to only one detector ($\mathbf{L}$ or $\mathbf{R}$) clicking is
\begin{equation}
	\begin{aligned}	q_{k_ak_b}^{L\theta}=&y_{k_ak_b}\left[e^{\omega_{k_ak_b}\cos\theta}-y_{k_ak_b}\right],\\
		q_{k_ak_b}^{R\theta}=&y_{k_ak_b}\left[e^{-\omega_{k_ak_b}\cos\theta}-y_{k_ak_b}\right].\\
	\end{aligned}
\end{equation}
where $\eta_a=\eta_d10^{-\alpha l_a/10}$, $\eta_b=\eta_d10^{-\alpha l_b/10}$, $y_{k_ak_b}=e^{\frac{-(\eta_ak_a+\eta_bk_b)}{2}}(1-p_d)$, and $\omega_{k_ak_b}=\sqrt{\eta_ak_a\eta_bk_b}$. The overall gain can be expressed as $q_{k_ak_b}= 1/2\pi\int_{0}^{2\pi} (q_{k_ak_b}^{L\theta}+q_{k_ak_b}^{R\theta})d\theta= 2y_{k_ak_b}[I_0(\omega_{k_ak_b})-y_{k_ak_b}]$, where $I_0(x)$ refers to zero-order modified Bessel functions of the first kind. The total number for $\{k_a,k_b\}$ is $x_{k_ak_b}=N p_{k_a}p_{k_b}q_{k_ak_b}$.

The post-matching events in the basis of $Z$ can be divided into two types: correct events $\{\mu_a \mathbf{o}_{a},\mathbf{o}_{b}\mu_b\}$, $\{\mathbf{o}_{a}\mu_a,\mu_b \mathbf{o}_{b}\}$,  and incorrect events $\{\mu_a\mathbf{o}_{a},\mu_b\mathbf{o}_{b}\}$, $\{\mathbf{o}_{a}\mu_a,\mathbf{o}_{b}\mu_b\}$. The corresponding numbers are denoted as
$n_C^z$ and $n_E^z$, respectively, which can be written as
\begin{equation}
	\begin{aligned}
		n_C^z = &x_{\min}\times\frac{p_{\mathbf{o}_a}p_{\mu_b}q_{\mathbf{o}_a\mu_b}}{p_{\mathbf{o}_a}p_{\mathbf{o}_b}q_{\mathbf{o}_a\mathbf{o}_b}+p_{\mathbf{o}_a}p_{\mu_b}q_{\mathbf{o}_a\mu_b}}\times\\
&\frac{p_{\mu_a}p_{\mathbf{o}_b}q_{\mu_a\mathbf{o}_b}}{p_{\mu_a}p_{\mathbf{o}_b}q_{\mu_a\mathbf{o}_b}+p_{\mu_a}p_{\mu_b}q_{\mu_a\mu_b}}
	\end{aligned} 
\end{equation}
and 
\begin{equation}
	\begin{aligned}
		n_E^z = &x_{\min}\times\frac{p_{\mathbf{o}_a}p_{\mathbf{o}_b}q_{\mathbf{o}_a\mathbf{o}_b}}{p_{\mathbf{o}_a}p_{\mathbf{o}_b}q_{\mathbf{o}_a\mathbf{o}_b}+p_{\mathbf{o}_a}p_{\mu_b}q_{\mathbf{o}_a\mu_b}}\times\\
  &\frac{p_{\mu_a}p_{\mu_b}q_{\mu_a\mu_b}}{p_{\mu_a}p_{\mathbf{o}_b}q_{\mu_a\mathbf{o}_b}+p_{\mu_a}p_{\mu_b}q_{\mu_a\mu_b}}.
	\end{aligned}
\end{equation}
The overall number of events in the $Z$ basis is $n^z= n^z_C+ n^z_E$.
Considering the misalignment error $e_d^z$,  the number of bit errors in the $Z$ basis is $m^z=(1-e_d^z)n^z_E + e_d^z n^z_C$, and the bit error rate in the $Z$ basis is $E^z=m^z/n^z$.

The overall number of events in the $X$ basis is 
\begin{equation}
	\begin{aligned}
		n^x=&4\int_\sigma^{\sigma+\delta} \frac{N  p_{\nu_a}p_{\nu_b}q_{\nu_a\nu_b}^{\theta}}{4\pi} d\theta\\
		=&\frac{Np_{\nu_{a}}p_{\nu_{b}}}{\pi}\times\\
  &\int_\sigma^{\sigma+\delta} y_{\nu_a\nu_b}(e^{\omega_{\nu_a\nu_b}\cos\theta}
		+e^{-\omega_{\nu_a\nu_b}\cos\theta}-2y_{\nu_a\nu_b})d\theta.\\
	\end{aligned}
\end{equation}
For simplicity, we only consider the case in which all matched events satisfy $\theta^i -\theta^j = 0$. In this case, when $r_a^i \oplus r_a^j \oplus r_b^i \oplus r_b^j=0$ (1), the $\{\nu_a^i\nu_a^j,~\nu_b^i\nu_b^j\}$ event is considered to be an error event when different detectors (the same detector) click at time bins $i$ and $j$.
The overall error count in the $X$ basis can be given as
\begin{equation}
	\begin{aligned}
		m^{x}=&4\int_\sigma^{\sigma+\delta}\frac{N  p_{\nu_a}p_{\nu_b}q_{\nu_a\nu_b}^{\theta}}{4\pi}  \times p_{E}d\theta\\
		=&\frac{2Np_{\nu_{a}}p_{\nu_{b}}}{\pi}\int_\sigma^{\sigma+\delta} y_{\nu_a\nu_b}\times\\
  &\left[\frac{(1-y_{\nu_a\nu_b})^{2}}{e^{\omega_{\nu_a\nu_b}\cos\theta}+e^{-\omega_{\nu_a\nu_b}\cos\theta}-2y_{\nu_a\nu_b}}-1\right]d\theta,\\
	\end{aligned}
\end{equation}		
where $p_{E}= \frac{2q_{\nu_a\nu_b}^{L\theta}q_{\nu_a\nu_b}^{R\theta}}{q_{\nu_a\nu_b}^{\theta}q_{\nu_a\nu_b}^{\theta}}$.  The code of our protocol has been uploaded to the open-source code website~\cite{xiecode}.

As a comparison, the overall error count in the $X$ basis of the AOPP is defined as
\begin{equation}
	\begin{aligned}
		\hat{m}^{x}&=4\int_\sigma^{\sigma+\delta} \frac{N  p_{\nu_a}p_{\nu_b}q_{\nu_a\nu_b}^{\theta}}{2\pi}\times \hat{p}_{E}d\theta\\
		&=\frac{2Np_{\nu_{a}}p_{\nu_{b}}}{\pi}\int_\sigma^{\sigma+\delta} y_{\nu_a\nu_b} (e^{-\omega_{\nu_a\nu_b}\cos\theta}-y_{\nu_a\nu_b})d\theta,\\
	\end{aligned}
\end{equation}
where we have $\hat{p}_{E}= \frac{q_{\nu_a\nu_b}^{R\theta}}{q_{\nu_a\nu_b}^{\theta}}$.


%

\end{document}